\documentclass[aps,prb,twocolumn,amsmath,amssymb,groupedaddress,10pt]{revtex4-1}
\pdfoutput=1

\usepackage{float}
\usepackage{graphicx}  % needed for figures
\usepackage{xcolor}
\usepackage{dcolumn}   % needed for some tables
\usepackage{bm}        % for math
\usepackage{amssymb}   % for math
\usepackage{epstopdf}
\usepackage{xfrac}

\usepackage{amsmath}
\usepackage{amsfonts}
\definecolor{darkblue}{rgb}{0.1,0.2,0.6}
\definecolor{darkred}{rgb}{0.8,0.1,0.2}
\usepackage[colorlinks,citecolor=darkblue,linkcolor=black,urlcolor=darkblue]{hyperref} 
\usepackage{tikz}
\usepackage{enumerate}
\usepackage{setspace}
\usepackage{url}  % This makes \url work

\begin{document}

%Title of paper
\title{Variational identification of a fractional Chern insulator in an extended Bose-Hubbard model}  
\author{Hassan Shapourian}
\author{Bryan K. Clark}
\affiliation{Department of Physics, University of Illinois at Urbana-Champaign, Urbana, Illinois 61801, USA}

\begin{abstract}

We study the extended Bose-Hubbard model on the square lattice at half filling as a function of next-nearest neighbor hopping amplitude and interaction strength. 
To variationally map out the phase diagram of this model, we develop a two-parameter family of wave-functions based on the parton construction which can describe both topological and broken symmetry phases on equal footing. In addition, our wave-functions resolve long standing issues with more conventional short-range Jastrow wave-functions.
Using this variational ansatz, we show that a spontaneous time-reversal symmetry breaking fractional Chern insulator is energetically favored over a critical region between two superfluid phases.
In verifying the properties of these parton wave-functions we exemplify a more robust way to identify topology through the Hall conductance. 

\end{abstract}

\maketitle

%%%%%%%%%%%%%%%%%%%%%%%%%%%%%%%%%%%%%%%%%%%%%%%%%%%%%%%%%
\textit{Introduction}--
Two broad categories of phases of matter are topological phases and spontaneous symmetry-breaking phases. 
The study of which microscopic Hamiltonians result in topological phases of matter has been the focus of significant recent work.  One robust way of finding topological phases is to find Hamiltonians where they are energetically preferred over all possible competing phases.  For example, in frustrated bosonic systems (equivalently frustrated magnets), one may wish to find systems where the fractional Chern insulator is energetically preferred to competing phases such as superfluids and Mott insulators.
 For small systems, exact methods such as exact diagonalization and the density-matrix renormalization group (DMRG) can be used~\cite{Misguich1999,Lhuillier2001,Mambrini2006,Lauchli2011, Messio2012,Yao2013,Bieri2015,White2011,Jiang2008,Depenbrock2012,Jiang2012,Vidal_CSL,Sheng_CSL,He2014}, but their exponential scaling (in either system size or width) limits their applicability to microscopic Hamiltonians with small correlation length.  In addition, DMRG can  get stuck in local minima further complicating the identification of the true phase.  Variational Monte Carlo (VMC)~\cite{Sorella2001,Sorella2003,Motrunich2005,Sorella2006,Lawler2008,Schroeter2007,Ran2007,*Hermele2008,Iqbal2011,*Iqbal2013,*Iqbal2014,Tay2011,Clark2011,Tao2012,Hu2013,Wang2013,Clark2013,Zhu14,Mei14}, while approximate, is a widely used alternative allowing for studies of large system sizes.  The quality of a variational study depends directly on the quality of the wave-functions being used to represent the respective phases.  These wave-functions must be efficient to evaluate and describe qualitatively the relevant phases at commensurate levels of accuracy. Otherwise, the better described phase may appear energetically superior over an artificially large range.  In addition, variational studies can be made immune to problems of local minima if the ansatz have only a few parameters.

\begin{figure}
\includegraphics[scale=0.48]{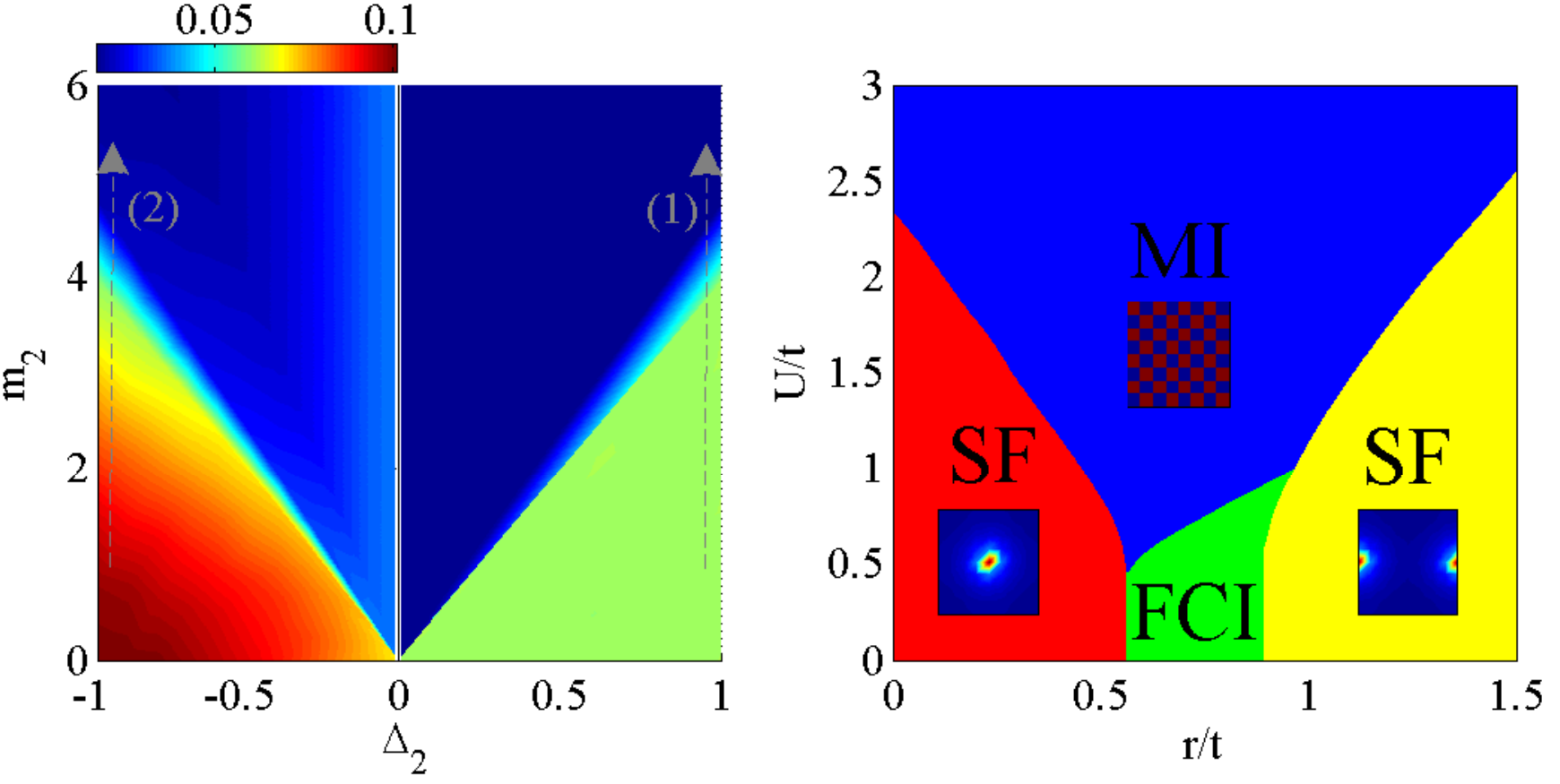}
\caption{\label{fig:int_phase} (Color online) Left: two-parameter space of parton wave-functions in terms of $f_2$ band parameters. In the left half, false color is the BEC fraction and in the right half, color represents the Chern number (blue=$0.0$, and green=$0.5$). We use a $40\times 40$ grid in $(m_2,\Delta_2)$ space. Right: variational phase diagram of the extended Bose-Hubbard model, Eq.~(\ref{eq:BH_model}). The inset on SF phases shows $\langle b_\textbf{k}^\dagger b_\textbf{k}\rangle$ where the hot spot is the condensate. The inset in MI shows CDW.}
\end{figure}

In this work, we use a two-parameter family of parton wave-functions to predict the existence of a fractional Chern insulator (FCI), as well as adjoining superfluid (SF) and Mott insulating (MI) phases (see Fig.~\ref{fig:int_phase}(right)), in an extended Bose Hubbard model in Eq.~(\ref{eq:BH_model}). Our class of wave-functions is efficient to evaluate and treats both topological and symmetry-breaking phases with one functional form minimizing the problem of non-uniform description quality.   The model we study is time-reversal symmetric and the observation of an emergent FCI phase in between two SFs is an example of spontaneous time-reversal symmetry breaking.   Given the developments of ultracold atomic gases in realizing a strongly interacting bosonic system subject to synthetic gauge fields~\cite{PhysRevLett.107.255301,*PhysRevLett.84.6,*PhysRevLett.87.120405,*Cooper_2008,*PhysRevLett.95.160404,*PhysRevLett.103.105303,*PhysRevLett.106.175301,*PhysRevLett.94.086803}, this model should be sufficiently simple to be realized experimentally.

The transitions between topological and broken symmetry phases have been studied using critical theories in terms of  Schwinger bosons, non-linear sigma models, as well as the Chern-Simons theories (anyon gas)~\cite{PhysRevLett.70.1501,Fisher_anyon_gas,PhysRevB.46.2223,Chubokov1994,Senthil_Motrunich,Isakov_PRB,IsakovScience,GroverSenthil2010,PhysRevB.86.075136,PhysRevLett.112.151601,Maissam2012,Maissam_Yao}.   
Our ansatz is constructed based on a parton decomposition motivated by field theoretic work of Barkeshli and McGreevy~\cite{Maissam2012}, 
where each boson is fractionalized into two fermions that are then glued back together to recover the physical Hilbert space. 
We use this fermionic parton construction, which should be considered as a microscopic construction for the Chern-Simons effective theory~\cite{PhysRevLett.70.1501,Fisher_anyon_gas,PhysRevB.46.2223},  to write the wave-functions in terms of products of determinants which can be efficiently evaluated at a polynomial cost in system size. 
Note that this efficient evaluation is not available in the bosonic constructions (e.g. Schwinger bosons), where the wave-function is written in terms of permanents %which limit the calculations to small system sizes 
whose evaluation is exponential in system size.  Working with wave-functions allows us to access physics beyond the mean-field which is particularly important because there are now classic examples where a parton description at mean-field does not accurately predict the underlying phase~\cite{TaoLi, Zhang_CSL_fail, PhysRevB.83.235122}.  
% In the parton construction studied here, each parton is described by a topological insulator with a topological invariant, i.e. Chern number, even in the case where the resulting phase is not topological. 
We note that the close parallel of our wave-functions to the field theory is valuable allowing for a connection between numerical work and analytic theory. %~\cite{PhysRevLett.70.1501,Fisher_anyon_gas,PhysRevB.46.2223,Maissam2012, Maissam_Yao,PhysRevB.86.075136}, 
It is interesting to note that our wave-functions resolve two standard problems with the more common short-range Jastrow used for bosonic phases by being size extensive and being able to have a zero condensate fraction.

To show our wave-functions accurately capture the different phases, we evaluate a series of observables (see Table~\ref{tab:tests}) on 1600 different wave-functions.  From Fig.~\ref{fig:int_phase} (left) we can see that the mapping from variational parameters to phases closely tracks the mean-field predictions. In the process of computing these observables, we exemplify that computing the topological degeneracy and Hall conductance is a more robust measure of the topological nature than the topological entanglement entropy.  In addition, we find the standard prescription for computing the Hall conductance must be supplemented to take into effect the relaxation of the internal gauge field which glues the partons together to get accurate results.  Having identified this mapping we are then able to  variationally identify which phase has the lowest energy throughout the parameter regime of our extended Bose-Hubbard model predicting the presence of a fractional Chern insulator.

\begin{table}
\begin{tabular}{ l  c c  c r } 
\hline
Observable & SF & MI & FCI & $\ \ $Fig. \\
  \hline                       
Momentum distribution $\langle n_{\textbf{k}}\rangle$  & BEC & -  & -  & (\ref{fig:int_phase},\ref{fig:single_den})\\
Static structure factor $\langle n_{\textbf{k}} n_{-\textbf{k}} \rangle$& - & CDW  & -  & (\ref{fig:Sq}) \\
Topological degeneracy on torus & - & 1  & 2  &(\ref{fig:svd_sym}) \\
Hall conductance $\sigma_{xy}$ & - & 0  & 1/2 & (\ref{fig:int_phase})  \\
\hline
\end{tabular}
\caption{\label{tab:tests} Four diagnostics to characterize the candidate wave-functions. The corresponding figures are indicated.}
\end{table}

%%%%%%%%%%%%%%%%%%%%%%%%%%%%%%%%%%%%%%%%%%%%%%%%%%%%%%%%%
%The Projective construction of bosonic wave-functions
\textit{Projective Construction}--
The projective construction approach has been used in the context of spin liquids\cite{Baskaran1987973,*Baskaran_PRB,*Affleck,Wen_book,fradkinbook}, the fractional quantum Hall (FQH)~\cite{Wen_1992,*PhysRevB.40.8079,*PhysRevLett.66.802,*PhysRevB.60.8827} and exciton bose liquids~\cite{PhysRevB.66.054526,*PhysRevB.75.235116,*PhysRevB.78.245111}. 
We study a system of hard core bosons on a square lattice at half filling.
Following the prescription of Ref.~\onlinecite{Maissam2012}, we define the fermionic representation of the charge one bosons by $b_i^\dagger = f^\dagger_{1,i} f^\dagger_{2,i}$
 where $f_{1,i}$ and $f_{2,i}$ correspond to two different flavors of charge one-half fermions and $i$ labels the lattice site. 
 This construction enlarges the Hilbert space and the constraint $b^\dagger_i b_i= f^\dagger_{1,i} f_{1,i}= f^\dagger_{2,i} f_{2,i} $ must be imposed. We consider the Hamiltonian that breaks the $SU(2)$ symmetry in the parton space to $U(1)$ (see Appendix~\ref{app:parton} for more details),
\begin{align} 
H=&  \sum_{\alpha, i,j}  t_{\alpha,ij} e^{i (A_{ij}/2-  q_\alpha a_{ij})} f_{\alpha,i}^\dagger f_{\alpha,j} \nonumber \\ 
&+ (A_0(i)/2 - q_\alpha a_0(i) ) \ \delta_{ij} f_{\alpha,i}^\dagger f_{\alpha,i} ,
\end{align}
where $A_\mu=(A_0,\vec{A})$ is the external gauge field, $\alpha=1,2$ labels the fermion flavor, and $q_{1(2)}=\pm1$ are the corresponding internal gauge charge of $f_{1(2)}$. The hopping amplitudes $t_{\alpha,ij}$ play the role of Hubbard Stratonovich fields which are static up to a fluctuating phase $a_{ij}$. The Lagrange multiplier $a_0(i)$ is introduced to ensure the particle number constraint. The combination of these fields $a_\mu=(a_0,\vec{a})$ therefore leads to an emergent  local $U(1)$ gauge symmetry.

\begin{table}
\begin{tabular}{ r  c  l } 
\hline
$f_2$ parameters \ \ \ \ $\ \ \  $ & $C_2\ \ \ $ & bosonic phase \\
  \hline                       
$\Delta_2>0, |m_2|< 4 |\Delta_2| \ \ \  $ & $ + 1\ \ \ $ & FCI $\nu=1/2$\\
$\Delta_2<0, |m_2|< 4 |\Delta_2| \ \ \  $ & $- 1\ \ \ $ & SF\\
$|m_2| >4 |\Delta_2| \ \ \  $ & $0\ \ \ $ & MI \\
  \hline  
\end{tabular}
\caption{\label{tab:phases} The bosonic phases in terms of the $f_2$ Chern number, $C_2,$ while the $f_1$ Chern number is fixed at $C_1=1$. The scale is set by  $t_2=t_1=2$.}
\end{table}

As a choice of $t_{\alpha,ij}$ parameters, we consider the $\pi$-flux square lattice model~\cite{Ludwig94} for both $f_1$ and $f_2$ fermions with three parameters for each (details in Appendix~\ref{app:pi-flux}): nearest neighbor hopping $t_{\alpha}$, next-nearest neighbor hopping $\Delta_{\alpha}$ and the on site mass term $m_{\alpha}$ where $\alpha=1,2$ indicates the corresponding parameter for $f_1$ and $f_2$ respectively. The spectrum of this model is gapped at half filling and the Chern number associated with the filled bands can be tuned by varying the above parameters. We fix the Hamiltonian for $f_1$:  $t_1=2\Delta_1=2$ and $m_1=0$ giving the lowest band unit Chern number $C_1=1$. Consequently, the phase diagram can be derived in terms of the $f_2$ parameters; i.e. $\Delta_2$ and $m_2$, see Fig.~\ref{fig:int_phase}(left) or Tab. \ref{tab:phases}.  
The bosonic wave-function is a projected product of two Slater determinants
\begin{align*}
\Psi (\{\textbf{r}_i\}) = \text{det}_1 (\{\textbf{r}_i\}) \times \text{det}_2(\{\textbf{r}_i\}),
\end{align*}
where the projection simply means the same configuration $\{\textbf{r}_i\}$ for both fermions.
Notice that for the FCI, these states are analogous to the composite fermion picture~\cite{PhysRevLett.63.199, Fisher_anyon_gas} used in FQH.
For the symmetry breaking phases, these wave-functions represent a novel alternative  to the more common approach, pioneered originally by Jastrow~\cite{Jastrow_55} and 
McMillan~\cite{McMillan}, of representing bosonic states as an expansion in short range pair product Jastrow factors; Jastrow wave-functions are constructed from classical observables making a universal scheme to incorporate fundamentally  topological quantum states difficult.  In addition, the parton framework for MI and SF naturally resolve two qualitative failures in short-range Jastrow's: the presence of a condensate fraction in the MI (i.e. forbidding supersolidity) and the lack of size consistency in the SF.  %These both form as the result of quantum or classical fluctuations. 
 Our parton wave-functions remove the condensate fraction naturally (e.g. Fig.~\ref{fig:pdist} in Appendix~\ref{app:SPDM}) and are straight-forwardly size consistent since a product of determinants can be written as a determinant of a single block diagonal matrix (explicit derivation in Appendix~\ref{app:szcons}).

\begin{figure}
\includegraphics[scale=0.7]{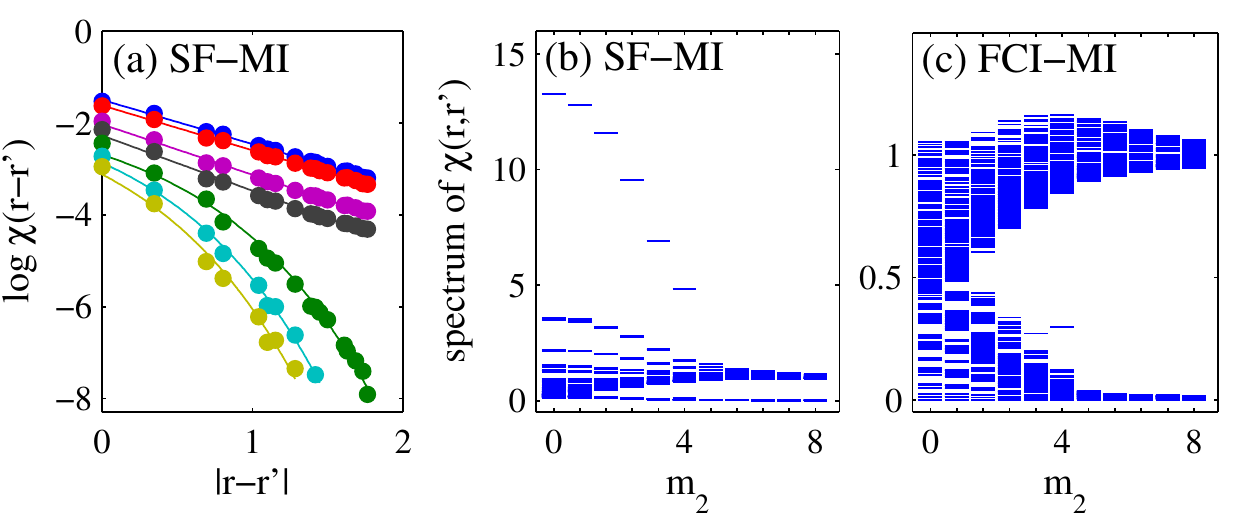}
\caption{\label{fig:single_den}  (Color online)   (a) SPDM $\chi(|\textbf{r}-\textbf{r}'|)$ for different values of $m_2$ along path(2) in Fig.~\ref{fig:int_phase}(left). Different colors from top to bottom represent $m_2=0, 1.6, 3.2, 3.9, 4.8, 6.4, 8$. Solid lines are fits. (b), (c) The entanglement spectrum of SPDM $\chi(r,r')$ as a function of $m_2$.  Errorbars are smaller than the symbols. The system size is $16\times 16$.}
\end{figure}

%%%%%%%%%%%%%%%%%%%%%%%%%%%%%%%%%%%%%%%%%%%%%%%%%%%%%%%%%
\textit{Mapping the wave-functions to different phases}-- 
We use variational Monte-Carlo to characterize the candidate wave-functions. 
Note that the wave-functions are constructed based on the $\pi$-flux Hamiltonian (Eq.~(\ref{eq:model}) in Appendix~\ref{app:pi-flux}) and they are complex-valued in general. Deep in the SF phase ($m_2=0$) the phases of the slater determinants exactly cancel each other and the resulting wave-function is real. This is indeed consistent with the flux smearing argument in the composite fermion picture~\cite{PhysRevLett.63.199, Fisher_anyon_gas}. As we deviate from $m_2=0$ towards the MI phase some configurations acquire a phase. However, we find that $\langle b_\textbf{r}^\dagger b_{\textbf{r}'}\rangle$, $\langle n_\textbf{r} n_{\textbf{r}'}\rangle$ and the variational energy only weakly depend on the phase angle of the wave-function and one can safely take the modulus or the real-part. We also note that the FCI wave-functions are equal to the SF wave-function times an extra phase angle which is responsible for the topological order. This is also a property of Laughlin wavefunctions~\cite{PhysRevLett.58.1252}.  
In the following, we compute a variety of observables as summarized in Table~\ref{tab:tests}.

\textit{Single-particle density matrix  (SPDM)}--
We first calculate the bulk  SPDM $\chi(\textbf{r},\textbf{r}')= \langle b_{\textbf{r}}^\dagger b_{\textbf{r}'} \rangle$ to search for off-diagonal long-range order. To see how SPDM behaves in different phases we take two paths in the phase diagram shown by the arrows in Fig.~\ref{fig:int_phase}(left).
We observe that in both FCI and MI phases SPDM always decays exponentially which means the quasi particle spectrum in these two states are gapped. However, in the SF phase the SPDM assumes a power-law $1/r^\alpha$ where $\alpha=0.95\pm 0.05$ consistent with $\alpha=1$ in the previous studies of the superfluidity in 2+1D bosonic systems~\cite{Fisher_orig_BH}. The transition from a power law behavior to an exponential decay occurs at the critical value $m_2=4\Delta_2$, at which $C_2$ changes from $-1$ to $0$ (see Fig.~\ref{fig:single_den}(a)). We represent the zero momentum (condensate) population $n_\textbf{0}=\langle b_\textbf{k}^\dagger b_\textbf{k} \rangle|_{\textbf{k}=0}$ by a color code to reconstruct the left half of the phase diagram in Fig.~\ref{fig:int_phase}(left). The agreement with the mean-field theory is quite remarkable.

We further use SPDM to establish the fact that the quasi particles in FCI and MI states are distinct in character: one is localized on the lattice sites (solid) and the other is delocalized over the lattice (fluid). 
To this end, we look at the entanglement spectrum of SPDM
\begin{align*}
\chi(\textbf{r},\textbf{r}') = \sum_\ell n_\ell \phi^*_\ell(\textbf{r}) \phi_\ell(\textbf{r}'),
\end{align*}
where $\phi_\ell(\textbf{r})$ (natural orbital) is the right eigenstate of SPDM with the eigenvalue $n_\ell$.
 This decomposition procedure is well-known as a conventional probe which identifies  a BEC when one of $n_\ell$'s is macroscopically occupied~\cite{Tony_book}. We verify this in the SF phase (see Fig.~\ref{fig:single_den}(b)). Deep in the MI, the spectrum is accumulated only at zero and one (i.e. particles fully localized on the lattice sites). In contrast, the spectrum in FCI is almost continuous between zero and one, showing the particle wave-functions are delocalized (Fig.~\ref{fig:single_den}(c)).

Moreover, we study the static structure factor and verify long-range order in the MI phase (Figs.~\ref{fig:Sq_low} and \ref{fig:Sq} in Appendix~\ref{app:SPDM}).

%\subsubsection{Response to external fields}

\begin{figure}
\includegraphics[scale=.7]{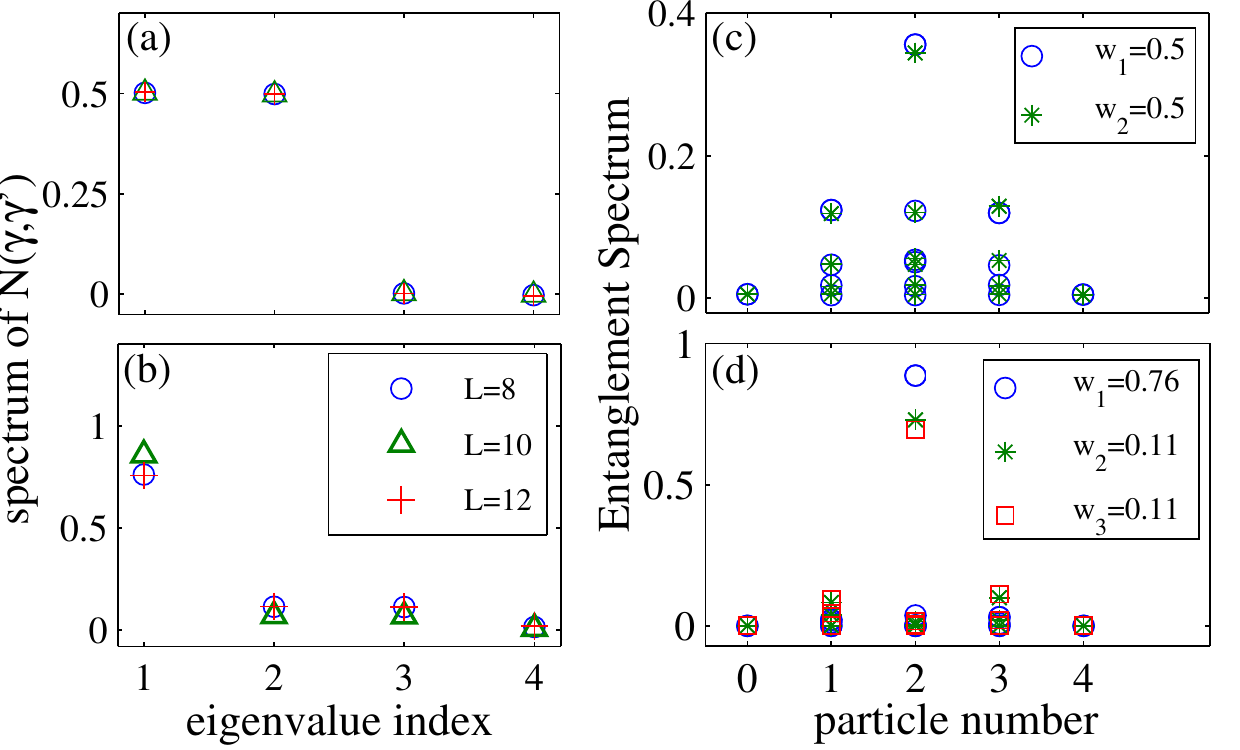}
\caption{\label{fig:svd_sym} The spectrum (weights) of the inner product matrix Eq.~(\ref{eq:inn_p_ext}) for FCI $(m_2=0)$ (a) and MI $(m_2=6)$ (b). (c) and (d) The eigenvalues of the full density matrix of a $2\times 2$ subsystem embedded in a $8\times 8$ system for the two highest weights in (a) [locally indistinguishable] and the three highest weights in (b) [locally distinguishable]. $\text{w}_i$ represents the weights.  Errorbars are smaller than the symbols.}
\end{figure}

\textit{Topological properties}--
For further wave-function characterization, we concentrate on the gapped phases and consider their topological properties. We calculate the topological degeneracy on a torus, Hall conductance and topological entanglement entropy (see Appendix~\ref{app:TEE}). However, we note that, because of finite size effects, the two former methods are much more accurate. To determine the degeneracy of the ground state, we study the change in the wave-function as a $2\pi$ flux quantum is inserted into either hole of the torus.   
We introduce the twisted boundary conditions associated with an external field
\begin{align}  \label{eq:twist}
\langle {\bf r}+ L_x \hat{x} | \Psi(\bar{\theta}_\gamma,\gamma) \rangle &= e^{i\gamma_x} \langle {\bf r}| \Psi(\bar{\theta}_\gamma,\gamma) \rangle, \nonumber \\
\langle {\bf r}+ L_y \hat{y} | \Psi (\bar{\theta}_\gamma,\gamma) \rangle &= e^{i\gamma_y} \langle {\bf r}| \Psi (\bar{\theta}_\gamma,\gamma) \rangle ,
\end{align}
where a bosonic many-body wave-function with an external twist angle $\gamma=(\gamma_x,\gamma_y)$ is denoted by $|\Psi(\bar{\theta}_\gamma,\gamma)\rangle$ and $L_x\times L_y$ is the system size. Note that the $\bar{\theta}$ is a static internal twist angle and can be found as the stationary solution of the effective action (details in Appendix~\ref{app:GSdeg})
which must be included in the parton Hamiltonian for computing the slater determinants; otherwise this method yields incorrect results. 
We compute the inner product matrix of four possible wave-functions
\begin{align} \label{eq:inn_p_ext}
N(\gamma,\gamma')= \langle \Psi(\bar{\theta}_\gamma,\gamma)|\Psi(\bar{\theta}_{\gamma'},\gamma') \rangle,
\end{align}
where $\gamma= (0,0), (0,2\pi), (2\pi,0)$ and $(2\pi,2\pi)$. The rank of this matrix yields the dimension of the ground state space. The normalized eigenvalues of $N(\gamma,\gamma')$ determine the weight of the corresponding eigenstate.  As shown in Fig.~\ref{fig:svd_sym}, there are two independent eigenstates with $0.5$ weight in FCI where the local density matrix is identical; in contrast, in the MI phase there is only one dominant eigenstate which implies that there is no degeneracy. By comparing the local density matrices, we also checked that the remaining weight in the eigenspectrum of the MI phase is locally distinguishable from the dominant eigenstate (see inset of Fig.~\ref{fig:svd_sym}(b)).
In Appendix~\ref{app:GSdeg}, we present an alternative method to compute the topological degeneracy purely based on the internal gauge field $a_\mu$. 

The Hall conductance of a many body wave-function is given in terms of the Chern number $C$, $\sigma_H= C e^2/h$~\cite{TKNN1982,Qi_MB_chern}.  We show that $C=0.50\pm 0.01$ per each state in FCI and $C=0.0\pm 0.01$ in MI. The Chern number of a non-degenerate many-body wave-function is computed in terms of an integration of the adiabatic curvature over the space of twisted boundary conditions (details in Appendix~\ref{app:Cnum}).

%\subsection{Phase diagram of the Bose-Hubbard model}

\textit{Microscopic Hamiltonian}-- Given these variational ansatz, one can ask if there is a microscopic Hamiltonian which supports the three phases. Considering the parton Hamiltonian as a mean-field ansatz for an interacting Hamiltonian we have found such a model which is basically an extension of the Bose-Hubbard model given by
\begin{align} \label{eq:BH_model}
H=&   -t \sum_{\langle i,j\rangle} (b^\dagger_i b_j + b^\dagger_j b_i) +r \sum_{\langle\langle i,j\rangle\rangle} (b^\dagger_i b_j + b^\dagger_j b_i) \nonumber \\ &+ U \sum_{\langle i,j\rangle}  n_i n_j
\end{align}
where $U$ is the nearest neighbor interaction and the hard-core limit is assumed. The nearest neighbor hopping $-t$ is negative while the next-nearest neighbor (diagonal) hopping $r$ is positive. This model can be viewed as a physical model (originally with all negative hopping amplitudes) subject to a $2\pi$ flux per plaquette. The large-$U$ limit favors the MI state. In the small-$U$ limit, $r \ll t$  gives SF with a condensate at $k=0$ and $r \gg t$ leads to SF with a condensate at $(0,\pi)$ or $(\pi,0)$. We find the FCI phase emerges between these two extreme limits.

To see this, we optimize the variational energy $E=\langle \Psi | H |\Psi \rangle/\langle \Psi | \Psi \rangle$ in $(m_2,\Delta_2)$-space of the trial wave-functions~\footnote{In constructing SF with a BEC at $\textbf{k}=(\pi,0)$, we use a slightly different model for partons. The noninteracting Hamiltonian has undergone a unitary transformation ($\pi/2$-rotation) in the sublattice space} for each value of $(r/t,U/t)$ and map out the phase diagram of Fig.~\ref{fig:int_phase}(right). 
The Hamiltonian of Eq.~(\ref{eq:BH_model}) can be mapped onto an extended version of the frustrated XXZ-model; in the spin language, the intermediate regime of $r/t$ will be a chiral-spin liquid.

As far as the transitions are concerned, we analyze the scaling behavior of the two-point correlation function $\langle b_{\textbf{r}}^\dagger b_{\textbf{r}'} \rangle$ and we find that SF-MI and SF-FCI transitions are continuous whereas the FCI-MI transition does not show a critical scaling behavior; in other words, the correlation length shows a discontinuity at the critical point instead of diverging to infinity. This behavior could be a signature of a first-order transition.  The critical coupling $t/U$ for the SF-MI transition along the vertical line $r=0$ line was found previously~\cite{XXZ_phase_2012,Batrouni_QMC_hardcore,Otterlo_BH} which agrees with our value of $(t/U)_c=2.2\pm 0.1$ in Fig.~\ref{fig:int_phase}(right).
We also compute the critical exponents and the anomalous dimension $\langle b_{\textbf{r}+\textbf{r}'}^\dagger b_{\textbf{r}'} \rangle \propto 1/r^{1+\eta}$ for the SF-MI and SF-FCI transitions (details in Appendix~\ref{app:crexp}) choosing as our variational single-particle orbitals the band structure at the mean-field transition.
Overall in both SF-MI and SF-FCI transitions, the evaluated anomalous dimension is rather large and in the case of SF-MI our result appears to be far from that of the 3D XY-model~\cite{PhysRevB.63.214503} while we get a consistent value of the critical exponent $\nu=2/3$.  This deviation is likely a failure of the variational ansatz to capture the critical point but it could also be finite size effects or a sign of a new fixed point~\cite{GroverSenthil2010,Isakov_PRB,IsakovScience}.

%%%%%%%%%%%%%%%%%%%%%%%%%%%%%%%%%%%%%%%%%%%%%%%%%%%%%%%%
%\section{Discussion}
In conclusion, we present a unifying scheme for constructing the candidate wave-functions to describe bosonic topological and non-topological phases. The variational parameters can  be tuned to transition from one phase to another.  We have introduced a simple microscopic Hamiltonian which variationally supports all the phases and allows for direct transitions between them; this Hamiltonian requires hopping (and optionally interactions) on a square lattice as well as gauge fields, both pieces have been already separately demonstrated in cold atom experiments. 

\textit{Acknowledgements}--
We would like to thank Maissam Barkeshli, Rebecca Flint, and David Pekker for insightful discussions and constructive comments on the manuscript. We would also thank Hitesh J Changlani and Abolhassan Vaezi for insightful discussions. Computation was done on Taub (UIUC NCSA). We acknowledge support from grant DOE, SciDAC FG02-12ER46875.

%%%%%%%%%%%%%%%%%%%%%%%%%%%%%%%%%%%%%%%%%%%%
\begin{appendix}
\section{\label{app:parton} Parton construction as a variational ansatz}
In this Appendix, we give a brief discussion of how the parton construction can be related to a microscopic interacting model. Consider the following generic physical Hamiltonian of bosons on a square lattice
\begin{align}
H= \sum_{i,j} [ t_{ij} e^{i A_{ij}} b^\dagger_i b_j - (\mu+A_0(i))\delta_{ij} n_i  + V_{ij} n_i n_j ]
\end{align}
where the unit charge bosons are coupled to an external gauge field ($A_\mu$) and $A_{ij}=\int_{r_i}^{r_j} A\cdot d\ell$. $t_{ij}$ and $V_{ij}$ define a hopping term and an interaction term between $i$ and $j$ sites respectively. The boson operator is decomposed into two parton fermions as
\begin{align}
b_i = f_{1,i} f_{2,i} = \frac{1}{2!} \epsilon_{\alpha\beta}\ f_{\alpha,i} f_{\beta,i} 
\end{align}
along with the constraint $n_{1,i}= f_{1,i}^\dagger f_{1,i} = f_{2,i}^\dagger f_{2,i}= n_{2,i}$.
Hence, the Hamiltonian is recast in the form
\begin{align}
H=& \sum_{i,j} ( t_{ij} e^{i A_{ij}}   f_{2,i}^\dagger f_{1,i}^\dagger f_{1,j} f_{2,j} + \text{H.c.})   + V_{ij} n_{2,i} n_{2,j}   \nonumber \\
&- \frac{(\mu+A_0(i))}{2}\delta_{ij} (n_{1,i }+n_{2,i})
\end{align}
where the interaction term is written only in terms of $f_2$ particles which is of course arbitrary. This construction possesses an internal $SU(2)$ symmetry in $(f_1,f_2)$ space; i.e. under a unitary transformation
\begin{align}
\tilde{f}_{\alpha',i} = U_{\alpha'\alpha} f_{\alpha,i}
\end{align}
where the boson operator transforms as
\begin{align}
\tilde{b}_i &=\frac{1}{2!} \epsilon_{\alpha'\beta'}\ f_{\alpha',i} f_{\beta',i} \nonumber \\
&= \frac{1}{2!}  \epsilon_{\alpha'\beta'} U_{\alpha'\alpha}  U_{\beta'\beta} f_{\alpha,i} f_{\beta,i } \nonumber \\
&= \det (U)\ b
\end{align}
the boson operator remains invariant as long as $\det(U)=1$; so the symmetry group is $SU(2)$. 
At this stage, we propose a mean-field ansatz 
\begin{align}
\chi_{1,ij} = e^{i A_{ij}/2} \langle f_{1,i}^\dagger f_{1,j} \rangle \ \ \ \ \ \  \ \chi_{2,ij} = e^{i A_{ij}/2} \langle f_{2,i}^\dagger f_{2,j} \rangle
\end{align}
which breaks the $SU(2)$ symmetry. Thus, the zeroth order Hamiltonian reads
\begin{align}
H =& \sum_{i,j} [t_{ij}  ( e^{i A_{ij}/2} \chi_{1,ij}  f_{2,i}^\dagger  f_{2,j}+ e^{i A_{ij}/2} \chi_{2,ij}  f_{1,i}^\dagger f_{1,j}  \nonumber \\
&-  \chi_{1,ij} \chi_{2,ij} )   -\frac{(\mu+A_0(i))}{2} \delta_{ij} (n_{1,i}+n_{2,i}) 
\nonumber \\  &+ V_{ij} (2\chi_{2,ii} n_{2,j} -\chi_{2,ii} \chi_{2,jj} ) ]  \nonumber \\  &+ \sum_i a_0(i) (f_{2,i}^\dagger f_{2,i} - f_{1,i}^\dagger f_{1,i})
\end{align}
where the static field $a_0(i)$ is introduced to impose the constraint
\begin{align}
\langle f_{2,i}^\dagger f_{2,i} \rangle =\langle f_{1,i}^\dagger f_{1,i} \rangle.
\end{align}
This Hamiltonian yields unphysical results for the following reasons:
first, the excitations are fermions which is inconsistent with the original bosonic model we start with; second, the Hilbert space is enlarged after fermions are introduced and the projection onto physical space is only performed at the mean-field level. Indeed, we must have $f_2^\dagger f_2 = f_1^\dagger f_1 $ at the operator level.

To resolve these issues we need to at least include the phase fluctuations~\cite{Wen_book}, $\chi_{\alpha,ij}=\bar{\chi}_{\alpha,ij} e^{i \text{sgn}(\alpha) a_{ij}}$
corresponding to the following $U(1)$ gauge transformation
\begin{align}
f_{1,i} &\to  e^{i\theta_i} f_{1,i}, \nonumber \\
f_{2,i} &\to  e^{-i\theta_i} f_{2,i},  \nonumber \\
a_{ij} &\to a_{ij} + \theta_i-\theta_j.
\end{align}
Therefore, the first order Hamiltonian would be
\begin{align} \label{eq:H_mf}
H=& \sum_{i,j} [t_{ij}  ( e^{i A_{ij}/2+i a_{ij}} \bar{\chi}_{1,ij}  f_{2,i}^\dagger  f_{2,j}  \nonumber \\  
&+ e^{i A_{ij}/2-i a_{ij}} \bar{\chi}_{2,ij}  f_{1,i}^\dagger f_{1,j}  \nonumber \\ &-  \bar{\chi}_{1,ij} \bar{\chi}_{2,ij} ) -\frac{(\mu+A_0(i))}{2} \delta_{ij} (n_{1,i}+n_{2,i}) \nonumber \\
&+ a_0(i) (n_{1,i}-n_{2,i})  +  V_{ij} (2\bar{\chi}_{2,ii} n_{2,j} -\bar{\chi}_{2,ii} \bar{\chi}_{2,jj} ) ].
\end{align}
Note that the $a_0(i)$ fluctuating field fixes the particle number and we recover our original physical Hilbert space. The above equation can be expressed as two separate non-interacting model for the fermions
\begin{align}
H_1=&  \sum_{i,j} [\tilde{t}_{1,ij} e^{i A_{ij}/2-i a_{ij}} f_{1,i}^\dagger f_{1,j}   \nonumber \\  &-(\frac{\mu+A_0(i)}{2} -a_0(i) ) \ \delta_{ij} n_{1,i} ], \\
H_2=& \sum_{i,j} [\tilde{t}_{2,ij}   e^{i A_{ij}/2+i a_{ij}}  f_{2,i}^\dagger  f_{2,j} + \tilde{V}_{2,i} n_{2,i} \delta_{ij}   \nonumber \\ 
&-(\frac{\mu+A_0(i)}{2} +a_0(i))\ \delta_{ij} n_{2,i}] ,
\end{align}
where new variables are defined by
\begin{align}
\tilde{t}_{\alpha,ij} &= t_{ij} \bar{\chi}_{\bar{\alpha},ij}, \\
\label{eq:int_des}
\tilde{V}_{\alpha,i} &= \sum_j V_{ij} \bar{\chi}_{\alpha,jj},
\end{align}
in which $\alpha=1,2$ and $\bar{\alpha}=2,1$. It is interesting to observe that the original hopping terms for bosons still remain as hopping terms for partons and the interaction terms appear in the form of a (on-site) mass term. The coefficients $t_{\alpha,ij}$ are chosen such that the spectra of the fermions are gapped at half filling where the lowest bands can have a non-zero Chern number. Under this condition, one can integrate out the fermionic fields and derive the low energy long-wavelength effective theory as
\begin{align} \label{eq:L_eff}
{\cal L} = \frac{\epsilon^{\mu\nu\lambda}}{4\pi} \sum_{\alpha=1,2}  C_\alpha  \left(A_\mu/2 - q_\alpha a_\mu\right) \partial_\nu \left(A_\lambda/2 - q_\alpha a_\lambda \right),
\end{align}
where $C_1$ and $C_2$ are the Chern numbers associated with $f_{1}$ and $f_2$ lowest bands respectively.  As was discussed in the main text, one can find different phases in the bosonic system (SF, MI, FCI) for various combinations of the Chern numbers (see Tab.~\ref{tab:tests} in the main text).

%%%%%%%%%%%%%%%%%%%%%%%%%%%%%%%%%%%%%%%%%%%%%%%%%%%%%%%%
\begin{figure}
\includegraphics[scale=1]{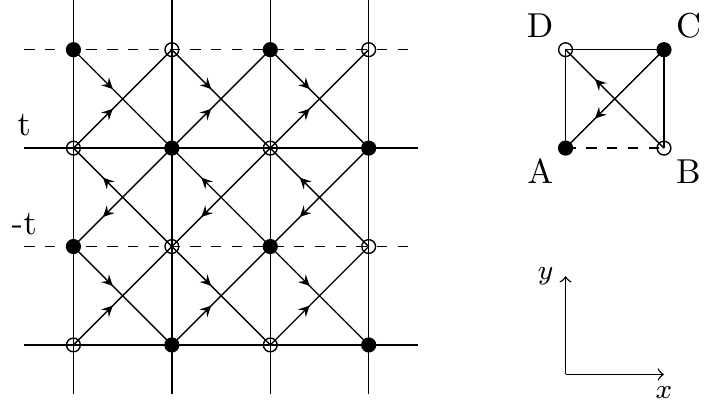}
\caption{\label{fig:lattice} Left: hopping matrix elements of the tight binding model in Eq.~(\ref{eq:model}).  Open circles and solid circles show positive($+m$) and negative($-m$) on-site energies respectively. Right: four-site unit cell with sublattice labeling.}
\end{figure}

\section{\label{app:pi-flux} Model Hamiltonian for Parton particles}
We consider the tight-binding Hamiltonian on a square lattice with a $\pi$-flux per plaquette
\begin{align}  \label{eq:model}
H= \sum_{\langle ij\rangle} t_{ij} c_i^\dagger c_j + i\sum_{\langle\langle ik\rangle\rangle} \Delta_{ik} c_i^\dagger c_k +  \sum_i m_i c_i^\dagger c_i 
\end{align}
where each site is defined by the coordinates $i=(i_x,i_y)$. This model is essentially a special case of the Hofstadter model with $\pi$-flux per plaquette and was originally used in Ref.~\onlinecite{Ludwig94} to study the integer quantum Hall transition. The hopping amplitudes obey the following rules
\begin{align} \label{eq:mf_param}
t_{i,i+\textbf{x}}&= (-1)^{i_y} t,   \nonumber \\
t_{i,i+\textbf{y}}&= t, \nonumber \\
\Delta_{i,i+\textbf{x}+\textbf{y}}&= -(-1)^{i_y} \Delta, \nonumber \\
\Delta_{i,i+\textbf{x}-\textbf{y}}&= (-1)^{i_y} \Delta, \nonumber \\
m_i &= (-1)^{i_x+i_y} m,
\end{align}
which is illustrated in Fig.~\ref{fig:lattice}. This is indeed equivalent to gauge choice of $A=\pi/a (-y/a,0)$ where $a$ is the lattice constant. The last (on-site) term is introduced to tune the model from the topological phase with non-zero Chern number to the trivial one.  According to our discussion in the previous Appendix (Eq.~(\ref{eq:int_des})), this term is a descendant of the boson interactions and its amplitude is directly related to the original interaction strength. For example, consider a next nearest neighbor interaction with strength $U$. An appropriate mean-field ansatz is the checkerboard density profile. Using the notations in the previous Appendix,
\begin{align}
\bar{\chi}_{1,ii}= (1+  (-1)^{i_x+i_y} )/2,
\end{align}
\begin{align}
\tilde{V}_{2,i}=\sum_j V_{ij} \bar{\chi}_{1,jj}= \frac{U}{2} \sum_{j \in \langle ij\rangle} \bar{\chi}_{1,jj} = U (1-  (-1)^{i_x+i_y} ).
\end{align}
The above ansatz is self-consistent in the parton basis and is equivalent to a checkerboard potential profile for parton fermions. After subtracting off the constant part, we can write it in the form of Eq.~(\ref{eq:mf_param}). This argument is general and can be applied to other forms of interactions. This point will be important if one wants to employ these wave functions as a variational guess for the ground state of a microscopic Hamiltonian. 

\begin{figure}
\includegraphics[scale=1]{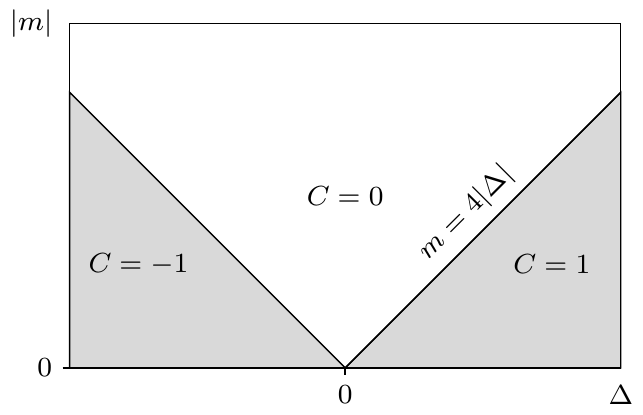}
\caption{\label{fig:nonint_phase} Chern number diagram of the parton Hamiltonian Eq.~(\ref{eq:model}) at half filling for $|\Delta|\leq |t|/2 $. The diagram is mirror symmetric with respect to the $m=0$ line.}
\end{figure}

We define a four-site unit cell (Fig.~\ref{fig:lattice}) and write the Hamiltonian in the momentum space in the form of $H=\sum_k c_k^\dagger h(k) c_k$ with
\begin{widetext}
%{\footnotesize \singlespacing
\begin{align} \label{eq:cellmodel}
h(k) = \left( \begin{array}{cccc}
m & -t (1+e^{ik_x})   & i\Delta  (e^{-i k_y}-1)(e^{i k_x}-1) &  t(1+e^{-i_ky})   \\
 -t (1+e^{-ik_x}) &  -m   & t(1+e^{-ik_y}) & -i\Delta  (e^{-i k_y}-1)(e^{-i k_x}-1)   \\
 -i\Delta (e^{i k_y}-1)(e^{-i k_x}-1) & t (1+e^{ik_y}) &  m & t(1+e^{-ik_x})   \\
    t(1+e^{i_ky}) & i\Delta   (e^{i k_y}-1)(e^{i k_x}-1) &t(1+e^{ik_x}) & -m 
\end{array}\right) ,
\end{align}
%}
\end{widetext}
which describes a four band model where multi-component fermion field is $c^\dagger= (c_A^\dagger,c_B^\dagger, c_C^\dagger, c_D^\dagger)$. 
In the regime $|\Delta|\leq |t|/2$ and at half filling where the lowest two bands are occupied, the effective long-wavelength Hamiltonian around the minimum of the conduction band $k=(\pi,\pi)$ reads
\begin{align}
h(k) &= \left( \begin{array}{cccc}
m & itk_x   & i4\Delta & it k_y \\
-itk_x &  -m &  itk_y & -i4\Delta   \\ 
-i4\Delta  & -i tk_y & m&  itk_x     \\
  -itk_y&  i4\Delta & -itk_x &  -m  
\end{array}\right) \nonumber \\
&= m \sigma_z -  tk_x \sigma_y - t k_y  \tau_y \sigma_x - 4\Delta \tau_y \sigma_z \nonumber \\
&= - t  (k_x \sigma_y  + k_y \tau_y \sigma_x) + (m - 4\Delta\tau_y) \sigma_z 
\end{align}
where $\sigma$ and $\tau$ are two sets of Pauli matrices. The eigenvalues of this Hamiltonian in $\tau$-space is simply $\tau_y =\pm 1$ which gives two copies of the 2+1 massive Dirac fields with opposite chiralities. As long as $|m|<4|\Delta|$ the Dirac points have opposite sign masses which gives a non-zero Chern number $C=\pm 1$. Once we pass the critical value of $m_c=4 |\Delta|$ both Dirac masses will have the same sign and we get the trivial insulator with $C=0$ for all values $|m|> m_c$. Figure~\ref{fig:nonint_phase} shows the phase diagram of this model. It is worth noting that on the horizontal line $m=0$ there is a direct transition from $C=+1$ to $C=-1$ where the critical theory is described by two massless Dirac fields.

%%%%%%%%%%%%%%%%%%%%%%%%%%%%%%%%%%%%%%%%%%%%%%%%%%%%%%
\begin{figure}
\includegraphics[scale=.6]{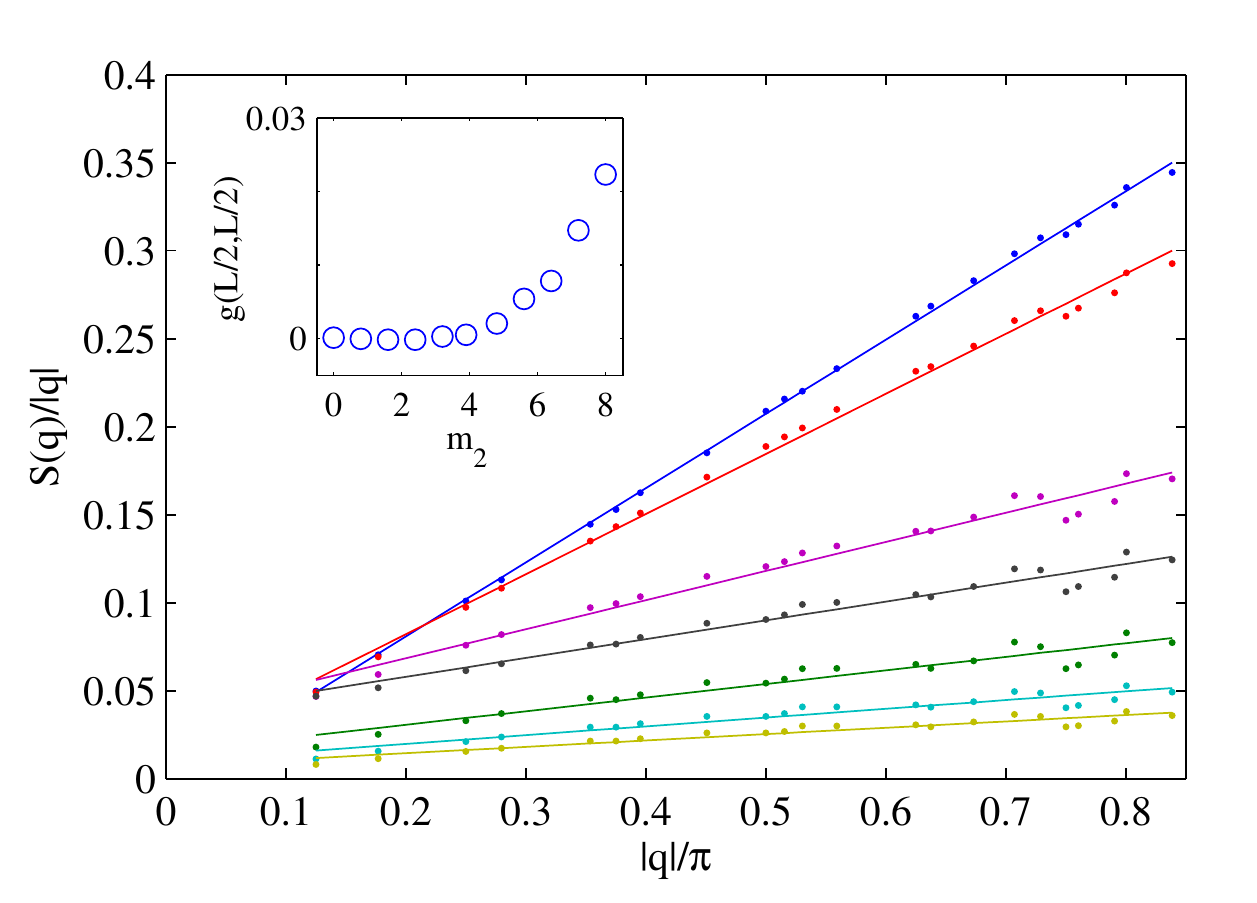}
\caption{\label{fig:Sq_low}  (Color online) The evolution of the structure factor per momentum $S(q)/|q|$ as the $f_2$ mass is tuned for the SF-MI or FCI-MI transition. The $f_2$ mass is changed from $m_2=0$ and $m_2=8$ as we move from the topmost curve to the lowest one in each panel. The jump in the slope corresponds to the transition point at $m_2=4$. The inset shows the emergence of the diagonal long-range range order in the MI phase in terms of the density-density correlation  $g(\textbf{r})$ at the long-distance limit .}
\end{figure}

\section{\label{app:SPDM} Single particle density matrix and structure factor}

\begin{figure}
\includegraphics[scale=0.55]{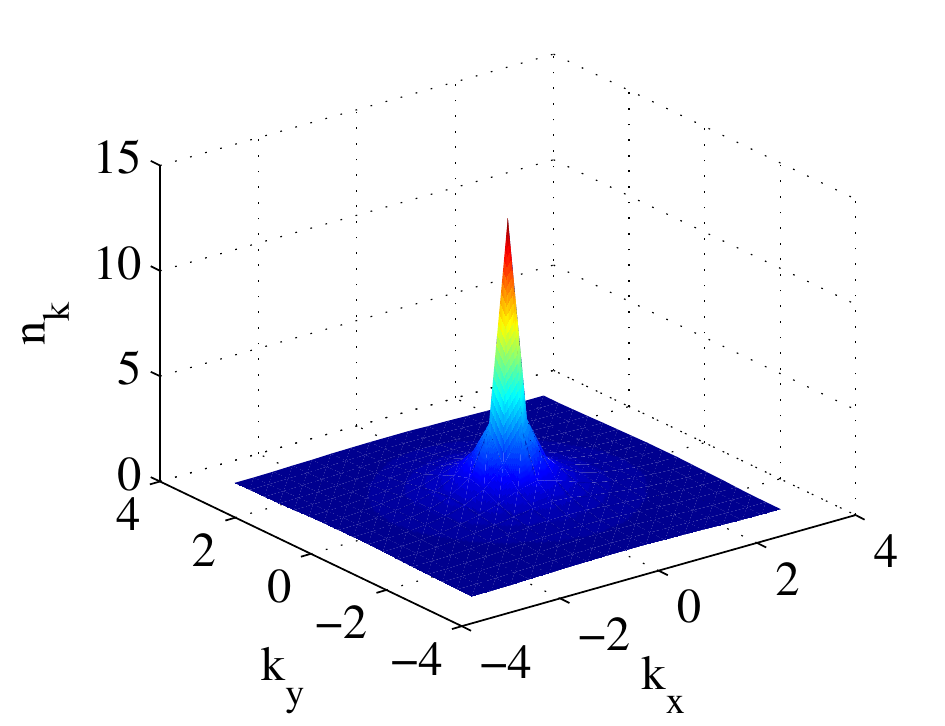}
\includegraphics[scale=0.55]{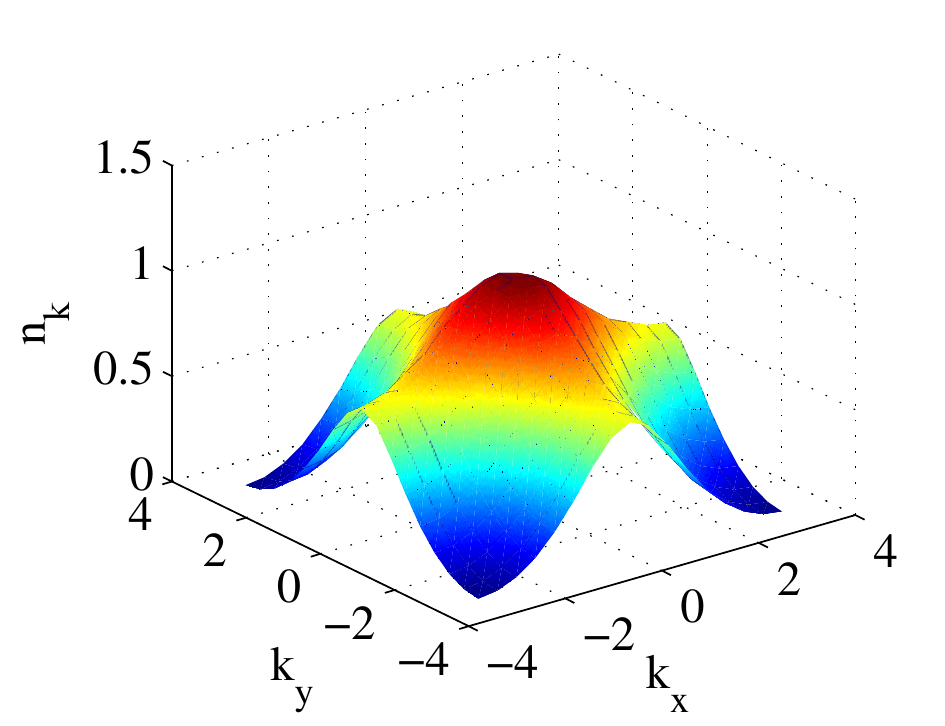}
\includegraphics[scale=0.55]{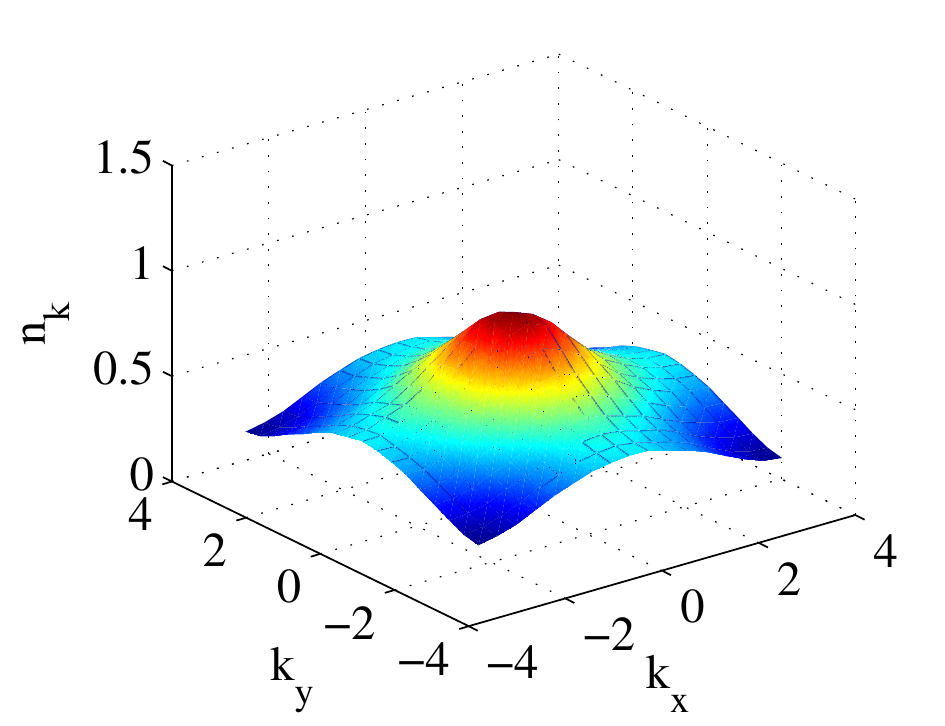}
\caption{\label{fig:pdist}  (Color online)  The momentum distribution $n_\textbf{k}$ as a function of lattice momentum $\textbf{k}=(k_x,k_y)$. From top to bottom: SF $(-1,0)$, FCI $(1,0)$, and MI $(1,8)$ where the pair of numbers are $f_2$ parameters $(\Delta_2,m_2)$.}
\end{figure}

The bulk single particle density matrix (SPDM) $\chi(\textbf{r},\textbf{r}')= \langle b_{\textbf{r}}^\dagger b_{\textbf{r}'} \rangle$ is shown in Fig.~\ref{fig:single_den}.
As we see, SPDM is exponentially decaying in both FCI and MI phases and there is a long-range order in SF.
The momentum distribution $n_\textbf{k}=\langle b_\textbf{k}^\dagger b_\textbf{k} \rangle$ associated with the population of the plane wave modes specified by the lattice momentum $\textbf{k}$ is plotted in Fig.~\ref{fig:pdist}. This quantity can be computed in terms of the 2D Fourier transform of the correlation function
\begin{align}
 \langle b_\textbf{k}^\dagger b_\textbf{k} \rangle=& \langle \left(\sum_\textbf{r} e^{i \textbf{k}\cdot \textbf{r}} b^\dagger_\textbf{r} \right) \left( \sum_\textbf{r'} e^{-i \textbf{k}\cdot \textbf{r}'} b_{\textbf{r}'} \right) \rangle \nonumber \\
 =& \sum_{\textbf{r},\textbf{r}'} e^{i \textbf{k}\cdot(\textbf{r}-\textbf{r}')}  \chi(\textbf{r},\textbf{r}').
\end{align}
The prominent peak at $\textbf{k}=0$ indicative of a BEC in the SF is evident. As we move towards the MI phase this peak becomes weaker and eventually disappears completely. In the MI, the momentum is almost distributed equally over all modes whereas in the FCI the population of modes slightly varies, particularly it vanishes at the corner points $\textbf{k}=(\pm\pi,\pm\pi)$. This could be interpreted as the FCI is completely featureless at lattice scales or there is no off-diagonal short range order in this fluid.

\iffalse
\begin{figure*}
\includegraphics[scale=0.7]{figs/pdist/spdm_comb.eps}
\caption{\label{fig:single_den} SPDM as $f_2$ mass is tuned from $m_2=0$ to $m_2=8$ [from top to bottom]. Left panel: SF-MI transition ($\Delta_2=-1$). Right panel: FCI-MI transition ($\Delta_2=1$).  Solid lines are fits to the data points. Colors represent the same value of $m_2$ in both graphs.}
\end{figure*}
\fi 

\begin{figure}
\includegraphics[scale=0.55]{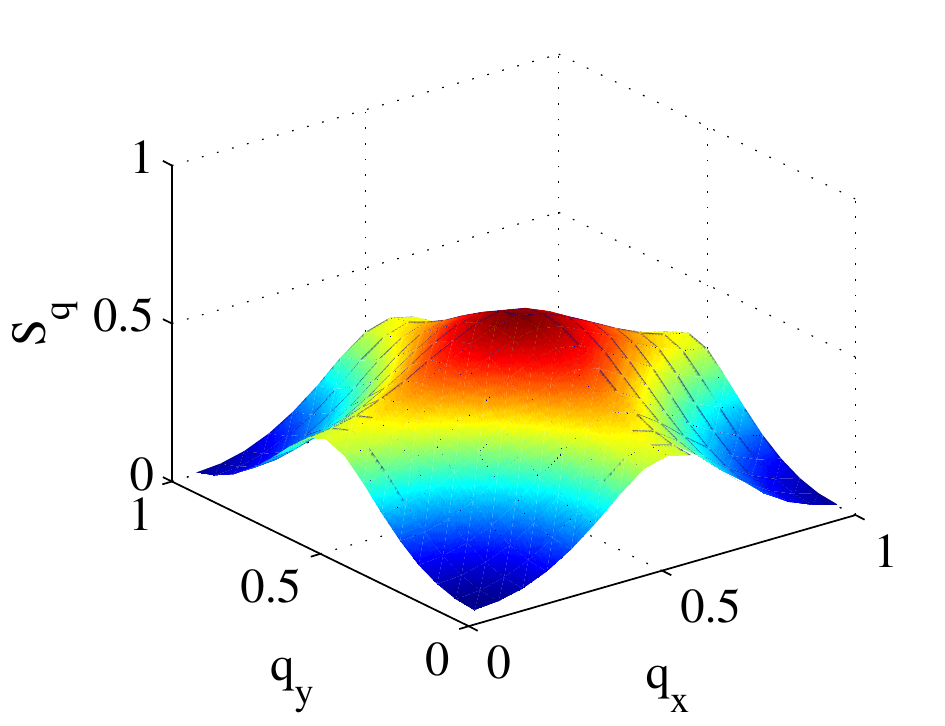}
\includegraphics[scale=0.55]{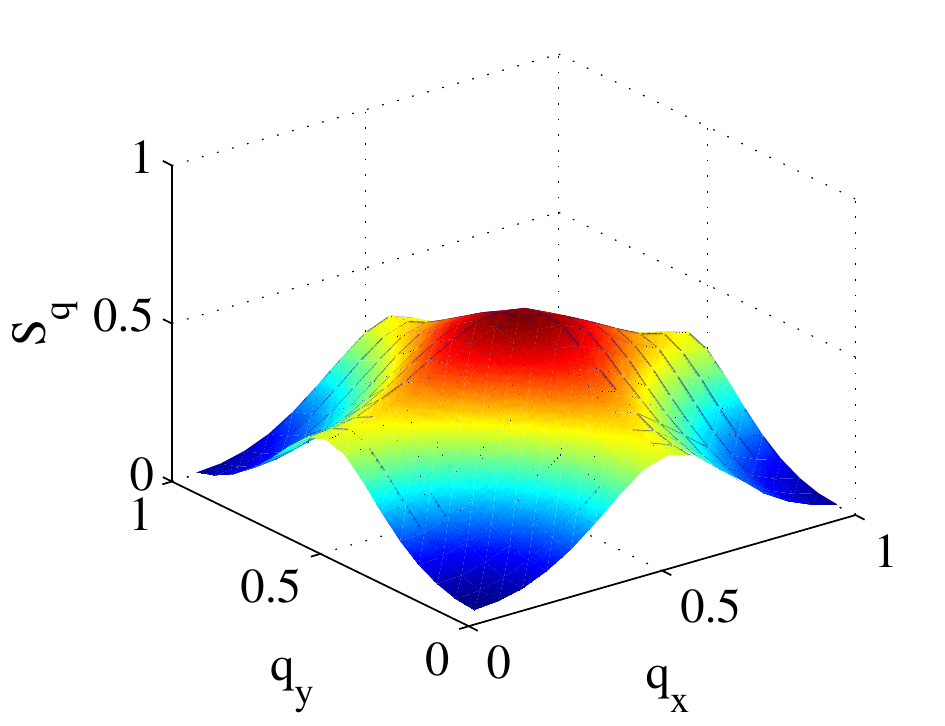}
\includegraphics[scale=0.55]{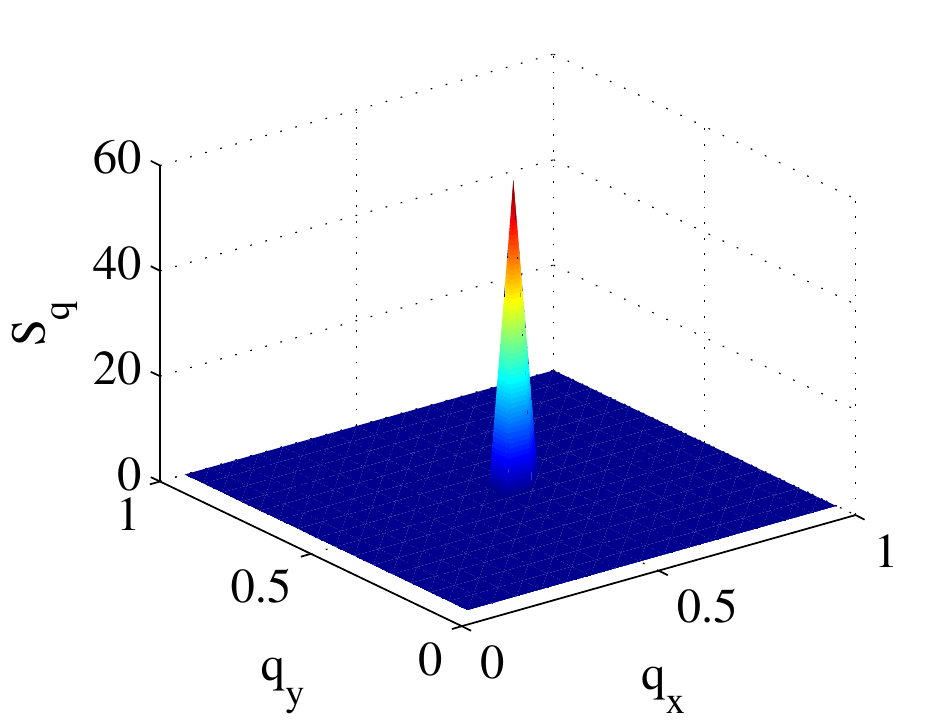}
\caption{\label{fig:Sq}  (Color online) The static structure factor $S(\textbf{q})$ as a function of lattice momentum $\textbf{q}=(q_x,q_y)$. From top to bottom: SF $(-1,0)$, FCI $(1,0)$, and MI $(1,8)$ where the pair of numbers are $f_2$ parameters $(\Delta_2,m_2)$.}
\end{figure}

To detect DLRO, we study the static structure factor which is defined by
\begin{align}
S(\textbf{q})= \langle n_\textbf{q} n_{-\textbf{q}}\rangle = \frac{1}{N} \sum_{\textbf{r},\textbf{r}'}  e^{-i \textbf{q}\cdot(\textbf{r} -\textbf{r}')}  \langle n_\textbf{r} n_{\textbf{r}'} \rangle,
\end{align}
where $N$ is the number of sites and $n_\textbf{r}= b^\dagger_\textbf{r} b_\textbf{r}$ is the particle number operator. As mentioned earlier, the wave-functions of SF and FCI are identical in modulus and therefore have the same structure factor. 
The evolution of the structure factor in the low-$q$ limit is illustrated in Fig.~\ref{fig:single_den}. There is a jump at the critical point ($m_2=4|\Delta_2|$) of the SF/FCI-MI transitions in accordance with the earlier results. Furthermore, there is a peak in the MI phase at $q=(\pi,\pi)$ corresponding to the checker board charge density wave (CDW) structure. This type of configuration is determined by the mass term which enlarges the unit cell in the parton Hamiltonian and should not be considered as a spontaneous breaking of the translational symmetry.  Recently, a different field theory~\cite{Maissam_Yao} that does incorporate spontaneous lattice translation symmetry breaking was developed.
 The inset of Fig.~(\ref{fig:Sq_low}) shows the long-distance real space density-density correlations, $g(\textbf{r}-\textbf{r}')=1- \langle n_\textbf{r} n_{\textbf{r}'}  \rangle / \langle n_\textbf{r} \rangle\langle n_{\textbf{r}'} \rangle$, for SF/FCI-MI transition. In the MI phase, the system develops a diagonal long-range order such that $\langle n_\textbf{r} n_{\textbf{r}'}  \rangle \neq \langle n_\textbf{r} \rangle\langle n_{\textbf{r}'} \rangle$.
The 2D structure factor $S(\textbf{q})$ for different phases is shown in Fig.~\ref{fig:Sq}. We have already noted that $S(\textbf{q})$ is identical for SF and FCI phases. 
The $(\pi,\pi)$ peak in MI phase corresponds to the checker board density structure. All the simulations in this Appendix are done for a $16 \times 16$ lattice consisting of $24 k$ sweeps over the lattice. Each data point is averaged over $18 k$ samples.

%%%%%%%%%%%%%%%%%%%%%%%%%%%%%%%%%%%%%%%%%%%%%%%%%%%%%%%%%
\begin{figure*}
\begin{tikzpicture}
\node[anchor=center,inner sep=0] at (0,0) {\includegraphics[scale=0.5]{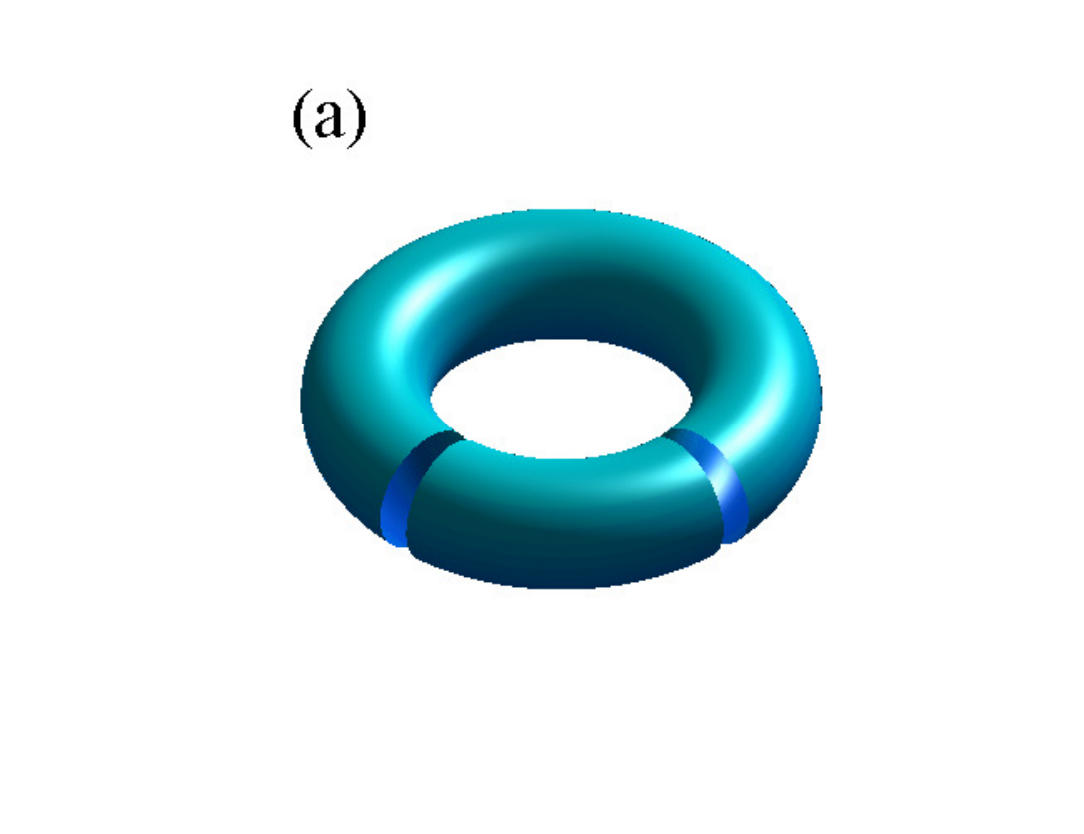}};
\draw[<->] (-0.6,-1.5) arc (250:290:2) ;
\draw[<->] (-.6,-0.8)  arc (160:180:2) ;
\draw (-0.7,-1.2) node[left] {\footnotesize $L_{Ax}$};
\draw (0,-1.7) node[below] {\footnotesize $L_{Ay}$};
\end{tikzpicture}
%\end{figure}
%\begin{figure*}
\includegraphics[scale=.65]{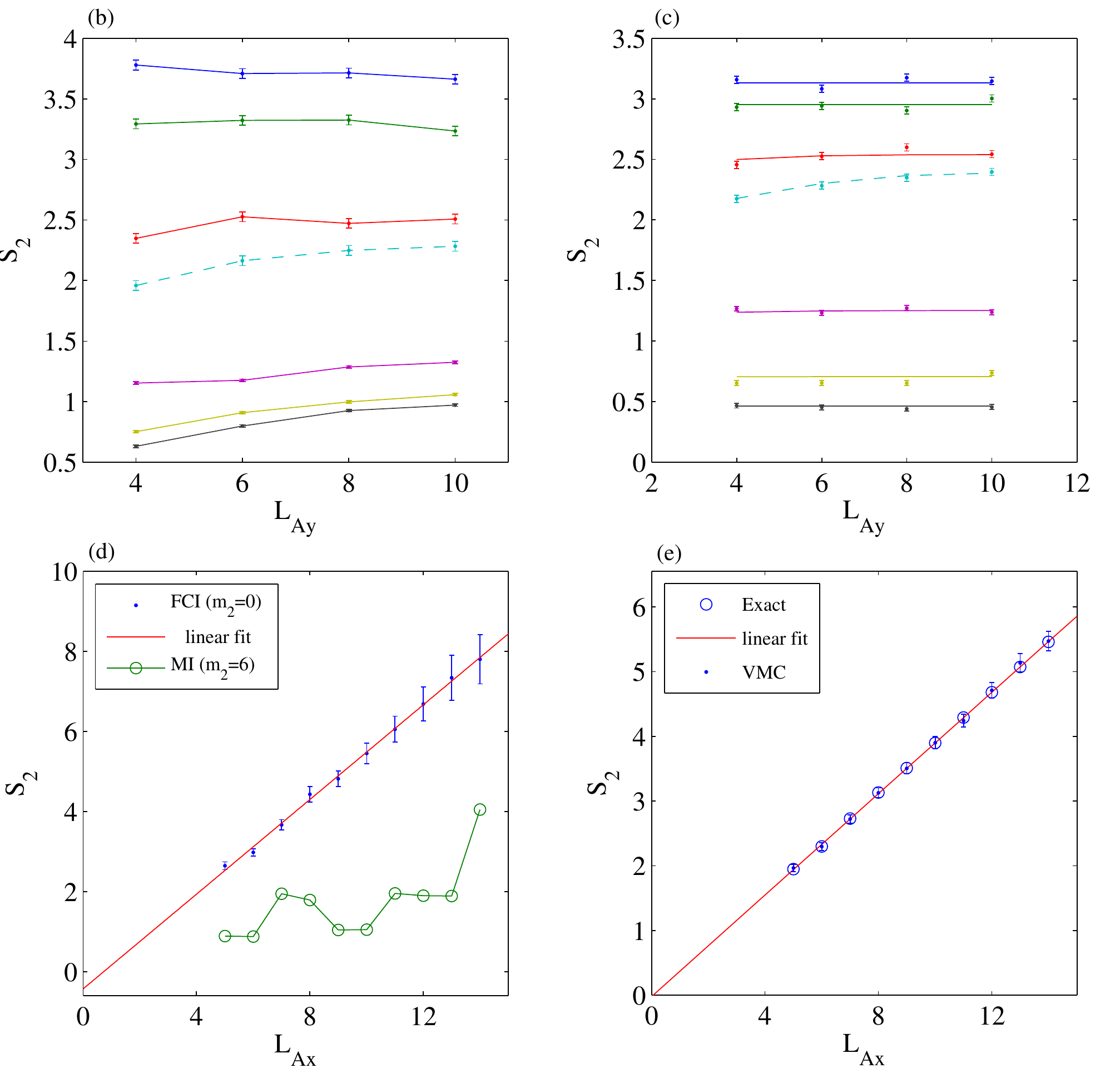}
\caption{\label{fig:EE}  (Color online) (a) The real-space cut. (b) and (c) REE as a function of subsystem longitudinal size $(L_{Ay})$. (b) FCI-MI transition. Solid lines are drawn as a guide to the eyes. (c) Non-interacting model Eq.~(\ref{eq:model}) with $\Delta=1$. Solid lines are the exact results. In (b) and (c), from top to bottom mass is varied $m_2=0, 1.6, 3.2, 4.0, 4.8, 6.4, 8.0$. Dashed lines correspond to the critical point $m_2=4$.
(d) and (e) REE as a function of subsystem lateral size $(L_{Ax})$.  (d) Interacting model where TEE for FCI is $\gamma=0.43\pm0.10$. Errorbars are smaller than symbol sizes for MI (green). (e) Non-interacting model Eq.~(\ref{eq:model}). TEE is $\gamma=0.02\pm0.02$ consistent with $\gamma=0$.}
\end{figure*}
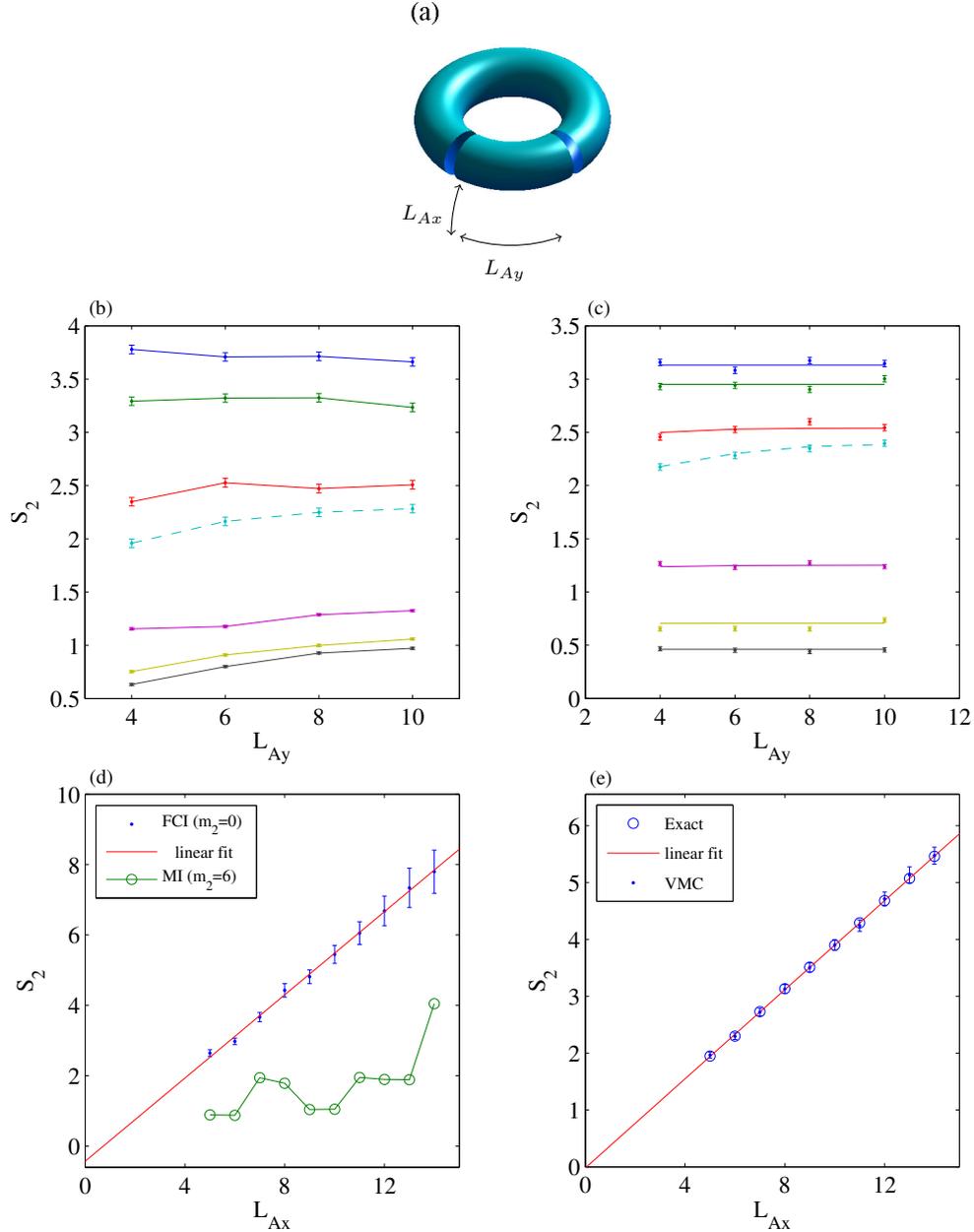

\section{\label{app:TEE} Entanglement entropy}
Here, we compute the second Renyi entanglement entropy (REE). We divide the system into two subsystems called $A$ and $B$. Using VMC, this is equivalent to calculating the expectation value of the SWAP operator~\cite{Zhang_CSL_fail,Ashvin2011}
\begin{align}
S_2 &= -\log \text{Tr} (\rho_A^2) = -\log \langle \widehat{\text{SWAP}} \rangle, \\
\langle \widehat{\text{SWAP}} \rangle &= \dfrac{\sum_{\substack{\alpha,\alpha' \\ \beta,\beta'}} \left|\Psi(\alpha,\beta)\right|^2 \left|\Psi(\alpha',\beta')\right|^2  \frac{\Psi(\alpha,\beta')\Psi(\alpha',\beta)}{\Psi(\alpha,\beta)\Psi(\alpha',\beta')}}{\sum_{\substack{\alpha,\alpha' \\ \beta,\beta'}} \left|\Psi(\alpha,\beta)\right|^2 \left|\Psi(\alpha',\beta')\right|^2},
\end{align}
where  $\alpha$, $\alpha'$ and $\beta$, $\beta'$ represent the configuration of the subsystems $A$ and $B$ respectively. Direct calculation of the SWAP operator from the above equation can give rise to large error bars, since the quantity $\Psi(\alpha,\beta')\Psi(\alpha',\beta)\Psi(\alpha,\beta)\Psi(\alpha',\beta')$ is zero most of the time when the number of particles in similar subsystems of the two wave functions are not equal. Following Ref.~\onlinecite{Zhang_CSL_fail}, to resolve this issue we break the above expression into a product of three quantities
\begin{align}
\langle \widehat{\text{SWAP}} \rangle &= \dfrac{\sum_{\substack{\alpha,\alpha' \\ \beta,\beta'}} \Psi^*(\alpha,\beta) \Psi^*(\alpha',\beta') \Psi(\alpha,\beta')\Psi(\alpha',\beta)}{\sum_{\substack{\alpha,\alpha' \\ \beta,\beta'}} \left|\Psi(\alpha,\beta)\right|^2 \left|\Psi(\alpha',\beta')\right|^2} \nonumber \\
&= \sum_{N_1} \langle S_{N_1,sgn} \rangle \langle S_{N_1,mag} \rangle  \langle \delta_{N_1,N_2} \rangle,
\end{align}
where
\begin{widetext}
\begin{align}
\langle S_{N_1,sgn} \rangle& = \dfrac{\sum_{\substack{\alpha,\alpha' \\ \beta,\beta'}} \left| \Psi^*(\alpha,\beta) \Psi^*(\alpha',\beta') \Psi(\alpha,\beta')\Psi(\alpha',\beta)\right| e^{i\theta(\alpha,\alpha',\beta,\beta')}}{\sum_{\substack{\alpha,\alpha' \\ \beta,\beta'}} \left| \Psi^*(\alpha,\beta) \Psi^*(\alpha',\beta') \Psi(\alpha,\beta')\Psi(\alpha',\beta)\right|} \Bigg|_{N_1=N_2}, \nonumber \\
\langle S_{N_1,mag} \rangle &= \dfrac{\sum_{\substack{\alpha,\alpha' \\ \beta,\beta'}} \left| \Psi^*(\alpha,\beta) \Psi^*(\alpha',\beta') \Psi(\alpha,\beta')\Psi(\alpha',\beta) \right|}{\sum_{\substack{\alpha,\alpha' \\ \beta,\beta'}} \left|\Psi(\alpha,\beta)\right|^2 \left|\Psi(\alpha',\beta')\right|^2} \Bigg|_{N_1=N_2} ,\nonumber \\
\end{align}
\end{widetext}
return the sign and the amplitude of the SWAP operator for a particular sector of having $N_1=N_2$ particles in $A$ subsystem. The third factor basically counts how often the condition $N_1=N_2$ is met. This decomposition is valid here since the SWAP operator is diagonal in the particle number basis. Moreover, this formula is quite efficient as we don't need to try all possible values for $N_1$ to obtain high accuracy results; in fact, we can easily first compute the frequency of different configurations $\langle \delta_{N_1,N_2} \rangle$ and find the dominant ones. In case of our wave functions, the configurations with half the number of sites being occupied is the most frequent and it drops to zero very fast as we go to the configurations with two or three more or less particles. Here, we calculate the contributions from the configurations with half filling plus/minus two particles.

We consider the full system on a $L_y \times L_x$ torus and cut a $L_{Ay}\times L_{Ax}$ piece in the middle of it (see Fig.~\ref{fig:EE}(a)). The REE as a function of $L_{Ay}$ (quasi-1D geometry) for FCI-MI transition is plotted in Fig.~\ref{fig:EE}(b). There is an apparent jump at the critical point. However, we cannot say much about the slope of REE vs $L_{Ay}$ due to finite size effects. We also vary $L_{Ax}$ where we expect REE follows the area law $S_2=\alpha L_{Ax} - \gamma$ where $\gamma$ is the topological entanglement entropy (TEE) (see Fig.~\ref{fig:EE}(d)). Note that the expected value of TEE for FCI is $\gamma=\log\sqrt{2}=0.3466$ and we have computed $0.43\pm 0.10$ which is consistent as was previously computed~\cite{Ashvin2011}. In the next Appendix, we use two other much more accurate methods to find the topological information. We will apply them to the entire phase diagram and find the phase boundary between FCI and MI (see Fig.~\ref{fig:int_phase}(left)). In addition, REE in MI phase shows an oscillatory behavior which means the effect of CDW at these system sizes is dominant. To extract TEE in this case, one needs to go to larger sizes and only compare system sizes that differ by multiples of $4$. To benchmark our calculations, we compare them with the exact results for the non-interacting model (Eq.~(\ref{eq:model})) in Figs.~\ref{fig:EE}(c) and (e). Figure~\ref{fig:EE}(c) shows that REE is not constant only at the transition point. Also, there is no TEE ($\gamma=0$) as shown in Fig.~\ref{fig:EE}(d).

%%%%%%%%%%%%%%%%%%%%%%%%%%%%%%%%%%%%%%%%%%%%%%%%%%%%%%%%%

\begin{figure*}
\includegraphics[scale=0.7]{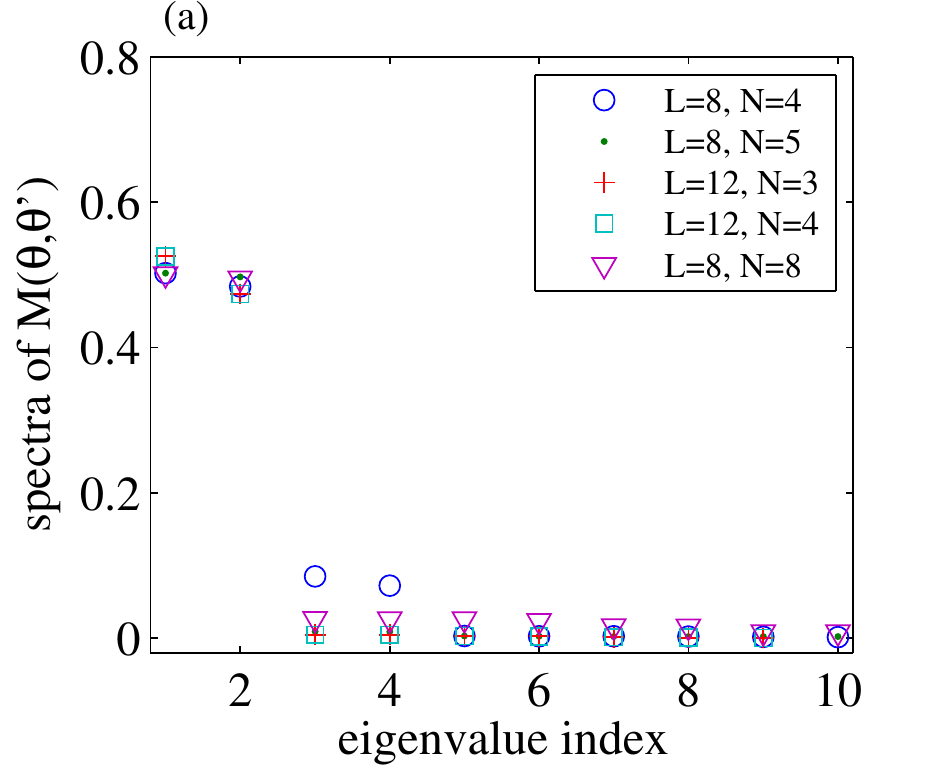}
\includegraphics[scale=0.7]{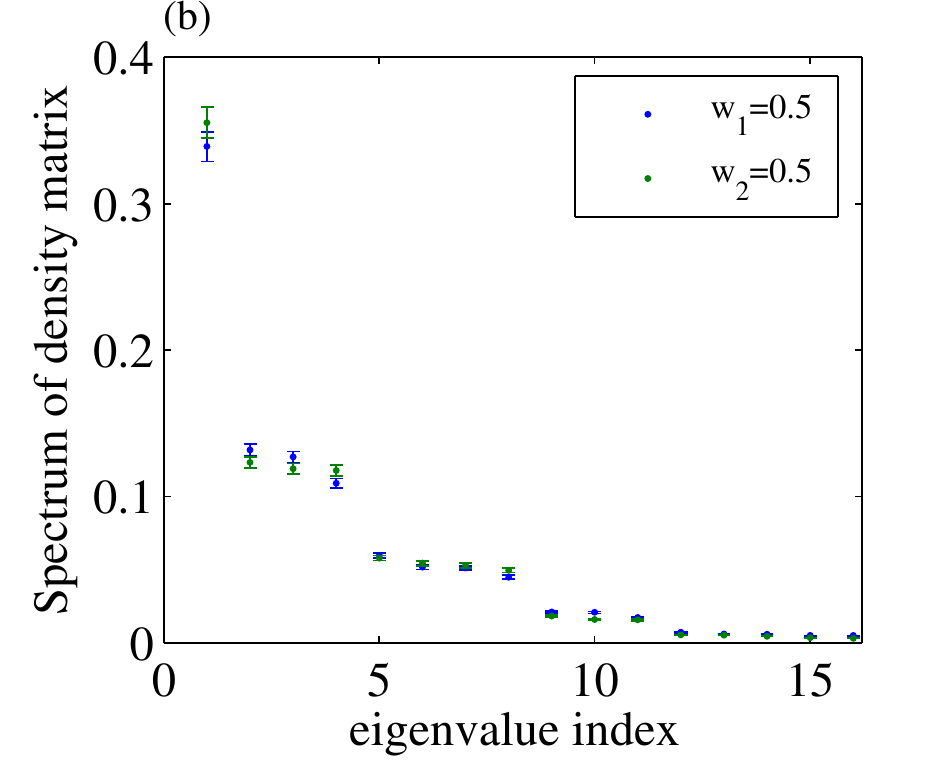}
\includegraphics[scale=0.7]{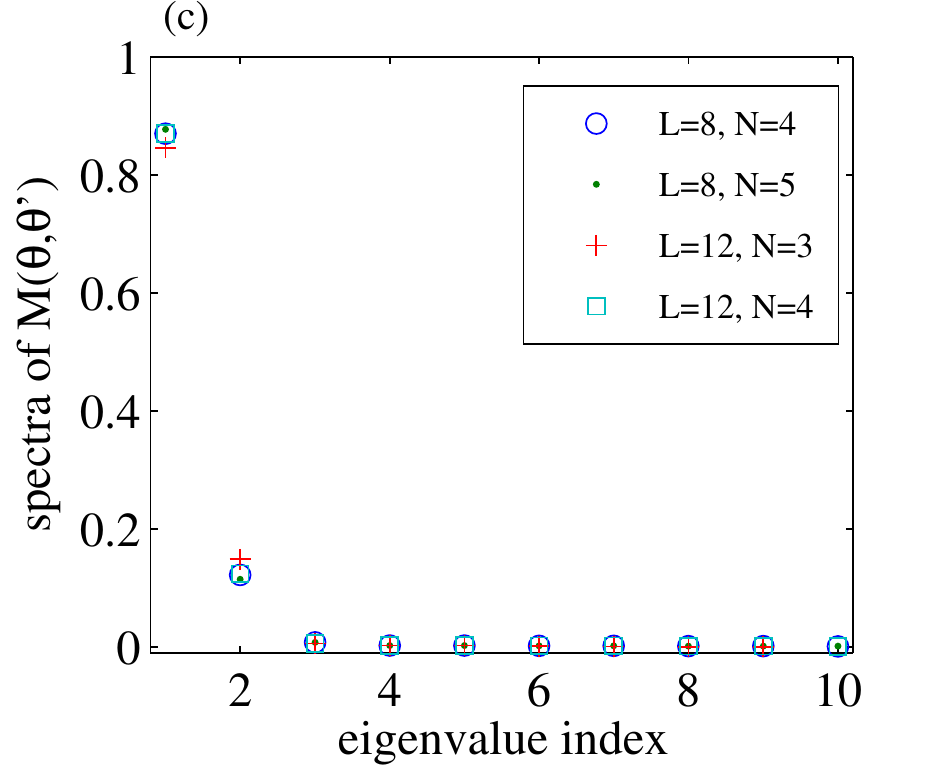}
\includegraphics[scale=0.7]{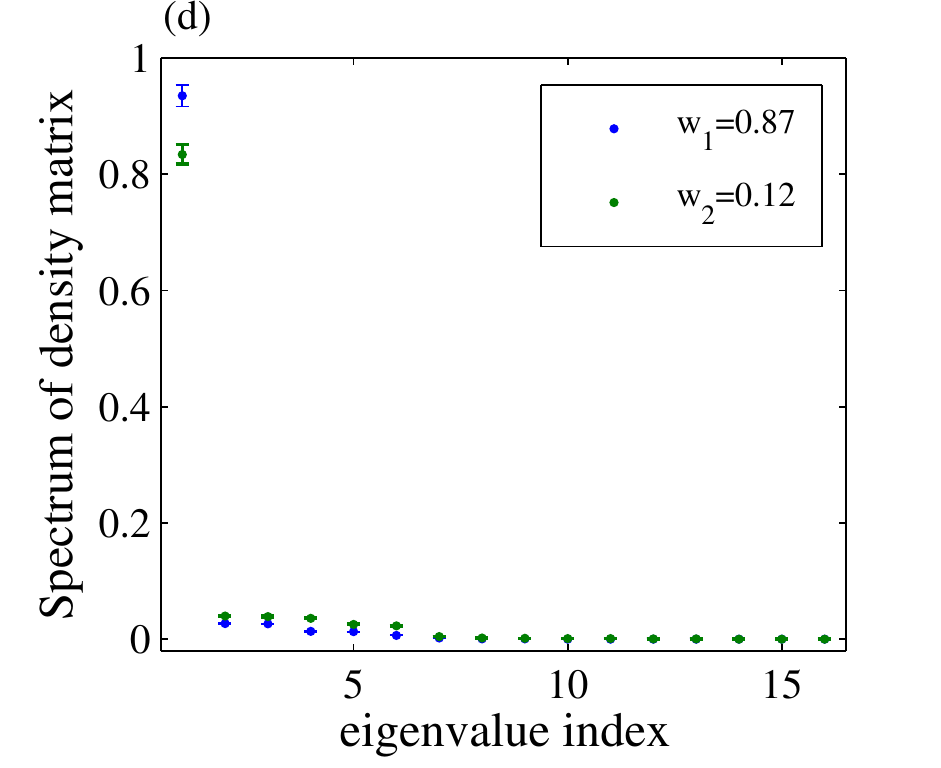}
\caption{\label{fig:svd_asym} The spectrum of the inner product matrix $M(\theta,\theta')$ Eq.~(\ref{eq:inn_p_int}) for FCI (a) and MI (c). The system size is $L\times L$ and the grid size in the internal gauge field space is $N\times N$. (b) and (d) show the eigenvalues of a full density matrix of a $2\times 2$ sub-system in a $8\times 8$ system. w represents the weight of $M(\theta,\theta')$ eigenstates.}
\end{figure*}

\section{\label{app:GSdeg}  Topological degeneracy using twisted boundary conditions}

We should emphasize that there are two types of twists in our model Eq. (\ref{eq:H_mf}): through the internal gauge field $a_\mu$ and through the external gauge field $A_\mu$. The former is an internal degree of freedom which can be treated as a parameter to span the ground state space. Here, we show that this space is two dimensional for the FCI consistent with the topological degeneracy of $\nu=1/2$ FCI on a torus. We have already shown the two-fold degeneracy of FCI by investigating the response to an external field and inserting a flux quantum into either holes of the torus. 

The projection is equivalent to fully integrating out the temporal component of the gauge field $a_0$ and the spatial component $\vec{a}$ is still left. We know that the excitations of $\vec{a}$ are gapped due to the existence of the Chern-Simons terms in the effective field theory at Eq.~(\ref{eq:L_eff}); however, we may still allow for the mean-field stationary solutions of $\vec{a}$. 
So, for every uniform configuration of $\vec{a}$ the projected wave-function is a ground state. If we denote the gauge field by $\vec{a}=(\theta_x/L_x,\theta_y/L_y)$ where $(L_x,L_y)$ are the dimensions of the system, it will describe the twisted boundary condition equal to $\pm(\theta_x,\theta_y)$ for $f_1$ and $f_2$ respectively.  
Recall that this is an internal phase acquired by the fermions with opposite signs and it will not appear for the bosonic particles simply because $b^\dagger=f_1^\dagger f_2^\dagger$.  

The variable $(\theta_x,\theta_y)$ forms a torus due to the $2\pi$ periodicity. To calculate the topological degeneracy in the ground state space, we look at the number of linearly independent wave functions. Numerically, the torus is discretized into a grid of points where the projected wave-function is denoted by $|\Psi(\theta)\rangle$ at the coordinate $\theta=(\theta_x,\theta_y) \in [0,2\pi) \times [0,2\pi)$. The inner-product matrix is constructed as 
\begin{align}\label{eq:inn_p_int}
M(\theta,\theta') = \langle \Psi(\theta)|\Psi(\theta') \rangle.
\end{align}
The rank of this matrix gives the dimension of the ground state space. In addition, one can look at the spectrum of $M(\theta,\theta')$ where the eigenstates associated with the non-zero eigenvalues form a basis in the ground state space. Moreover, the normalized eigenvalues determine the weight of the corresponding eigenstate. Figures \ref{fig:svd_asym}(a) and  \ref{fig:svd_asym}(c) show the spectrum of $M(\theta,\theta')$ for FCI and MI phases. The FCI phase has only two non-zero eigenvalues with equal weights. In MI, there is one dominant eigenvalue which span the $87\%$ of the space; however, there is another state which weighs around $12\%$. This should not be considered as a topological degeneracy. Indeed we have realized that these two eigenstates are locally distinct. More precisely, we calculate the eigenvalues of the full density matrix of a $2\times 2$ sub-system and show that they are different (see Fig.~\ref{fig:svd_asym} (b) and (d)). This means that these states are locally distinguishable. In Appendix~\ref{app:SPDM}, we observe that MI is described by a checker board charge density wave (CDW) order. Comparing these two wave functions reveals that one wave-function never reaches the full CDW order over the entire lattice during Monte Carlo runs as the wave function amplitude is extremely small for this configuration. This situation can also be viewed as the presence of a defect in CDW which prevents a full order.

\begin{figure*}
\begin{center}
\includegraphics[scale=.6]{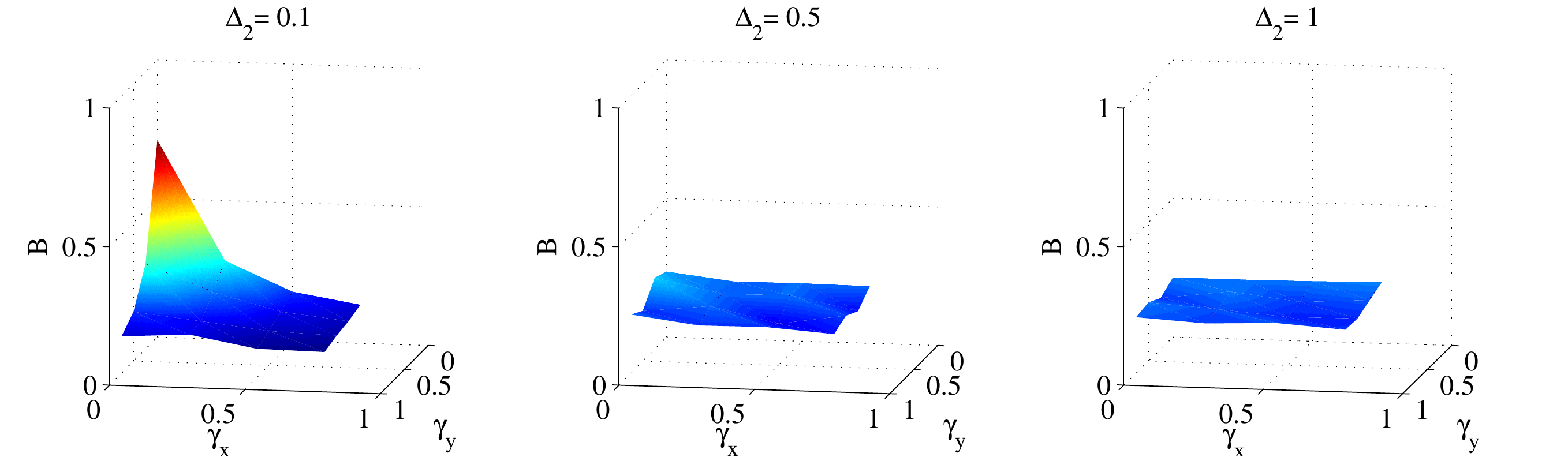}
\caption{\label{fig:Berry}  (Color online)  The Berry curvature in FCI phase for various values of $\Delta_2$. The system size is $8\times 8$ and the grid size in the twist-space is $5\times 5$.}
\end{center}
\end{figure*}

We shall now discuss how to incorporate the external boundary conditions. A twisted boundary condition is defined in Eq.~(\ref{eq:twist}) of the main text. It basically means that when a boson passes through the boundary, the wave function acquires an additional phase of $\gamma_x$ or $\gamma_y$ depending on which boundary is crossed. 
This boundary condition is gauge equivalent to the nonzero space components of the external gauge field $A_\mu=(A_0,\vec{A})$ introduced in Eq.~(\ref{eq:H_mf}); i.e. $\vec{A}=(\gamma_x/L_x,\gamma_y/L_y)$ where $L_x$ and $L_y$ are the dimensions of the system.
We should keep in mind that the slave particles are half-charged and a $\gamma$ phase twist for the bosons is equivalent to a $\gamma/2$ phase twist for the slave fermions with the same sign (see Eq.~(\ref{eq:H_mf})).
In order to determine the degeneracy of the ground state space, we compare the wave functions obtained in the presence or absence of a flux quantum threaded into either holes of the torus. If we na\"{i}vely impose the boundary conditions without considering the effect of the internal gauge field we will arrive at wrong results (see also Ref.~\onlinecite{Liu14} for discussion on a similar issue). 
 In fact, one must view the flux insertion as a dynamical process. For the purpose of studying the ground state,  this process has to be done in the adiabatic limit which lets the system relax to the instantaneous changes during the insertion. To the leading order and in the adiabatic limit where the rate of flux insertion is much smaller than the excitation gap of the system, the effective action would read
\begin{align}
{\cal L}_{eff}= \frac{1}{4\pi} \sum_{\alpha=1,2} C_\alpha  (\gamma_y/2 -q_\alpha \theta_y)\ \partial_t (\gamma_x/2 -q_\alpha \theta_x),
\end{align}
which is basically Eq.~(\ref{eq:L_eff}) where $\vec{a}(t)=(\theta_x/L_x,\theta_y/L_y)$. The stationary (saddle-point) solution $\frac{\partial {\cal L}_{eff}}{\partial \theta}=0$  is given by
\begin{align} \label{eq:sadd_pt}
\bar{\theta}= \left( \frac{C_1-C_2}{C_1+C_2}\right) \frac{\gamma}{2}.
\end{align}
This effect can be attributed as some sort of a screening~\cite{Liu14} or as a constraint to keep the total charge/current conserved~\cite{Mei14}.  So, for any choice of external fields, we must first compute the mean-field value and plug it into the Hamiltonian of Eq.~(\ref{eq:H_mf}) and then generate the wave functions. Therefore, the effective twist angles ($\tilde{\gamma}$) are found to be
\begin{align}
\tilde{\gamma}_1 &= \left( \frac{C_2}{C_1+C_2}\right) \gamma, \nonumber \\
\tilde{\gamma}_2 &=  \left( \frac{C_1}{C_1+C_2}\right) \gamma.
\end{align}
for $f_1$ and $f_2$ respectively. For instance, in FCI ($C_1=C_2=\pm1$) the overall phase $\tilde{\gamma}_1=\tilde{\gamma}_2=\gamma/2$ is purely due to the external fields and $\bar{\theta}=0$. However, in MI phase when $C_2=0$ we need to include a twist angle of $\gamma/2$ only for $f_2$ and the twist angle for $f_1$ vanishes.

%%%%%%%%%%%%%%%%%%%%%%%%%%%%%%%%%%%%%%%%%%%%%%%%%%%%%%%%%

\section{\label{app:Cnum} Method of computing Chern number}
The topological invariant (Chern) number of a non-degenerate many-body wave function can be computed in terms of an integration of the adiabatic curvature over the space of twisted boundary conditions ${\bf \gamma}=(\gamma_{x},\gamma_{x})$ which is a compact manifold (torus).
\begin{align}
C&=\frac{1}{2\pi i} \int d^2{\bf\gamma}\ \  \mathcal{B}(\gamma ).
\end{align}
The adiabatic curvature $\mathcal{B} = \partial_{x} {\cal A}_y -\partial_{y} {\cal A}_x$ is written in terms of the connection ${\cal} A_j(\gamma)= \langle \Psi (\bar{\theta}_\gamma,\gamma) |\partial_j |\Psi(\bar{\theta}_\gamma,\gamma)\rangle$. Numerically, we define a grid on the torus where points ${\bf \gamma}_\ell=(\gamma_{\ell x},\gamma_{\ell x})$ are separated by $\delta_x$ and $\delta_y$ along each direction. Therefore, the adiabatic curvature is evaluated by the following gauge-invariant expression
\begin{align} \label{eq:chern}
\mathcal{B}_\ell &= \log U_1({\bf \gamma}_\ell) U_2({\bf \gamma}_\ell+\delta_x)  U_1({\bf \gamma}_\ell+\delta_y)^{-1} U_2({\bf \gamma}_\ell)^{-1}
\end{align}
where 
\begin{widetext}
\begin{align}
U_\mu(\gamma_\ell)= \frac{\langle \Psi({\bar{\theta}_{\gamma_\ell},\gamma_\ell}) |\Psi({\bar{\theta}_{\gamma_\ell+\delta_\mu},{\gamma_\ell+\delta_\mu}})\rangle}{\sqrt{\langle \Psi({\bar{\theta}_{\gamma_\ell},\gamma_\ell}) | \Psi({\bar{\theta}_{\gamma_\ell},\gamma_\ell}) \rangle \langle \Psi({\bar{\theta}_{\gamma_\ell+\delta_\mu},{\gamma_\ell+\delta_\mu}})|\Psi({\bar{\theta}_{\gamma_\ell+\delta_\mu},{\gamma_\ell+\delta_\mu}})\rangle }}.
\end{align}
\end{widetext}
The Chern number is $C=\sum_\ell \mathcal{B}_\ell/2\pi$.  We perform these calculations on a $5\times 5$ grid in $\gamma$-space for $8\times 8$ and $10\times 10$ system sizes which yield the same value.
The Berry curvature is shown in Fig.~\ref{fig:Berry}. It is interesting to note that the Berry curvature is non-uniform in the vicinity of the critical point and this suggests that a simple effective description by the Chern-Simons action might not be valid.

%%%%%%%%%%%%%%%%%%%%%%%%%%%%%%%%%%%%%%%%%%%%%%%%%%%%%%%%%
\section{\label{app:crexp} Critical exponents}

\begin{figure*}
\includegraphics[scale=.65]{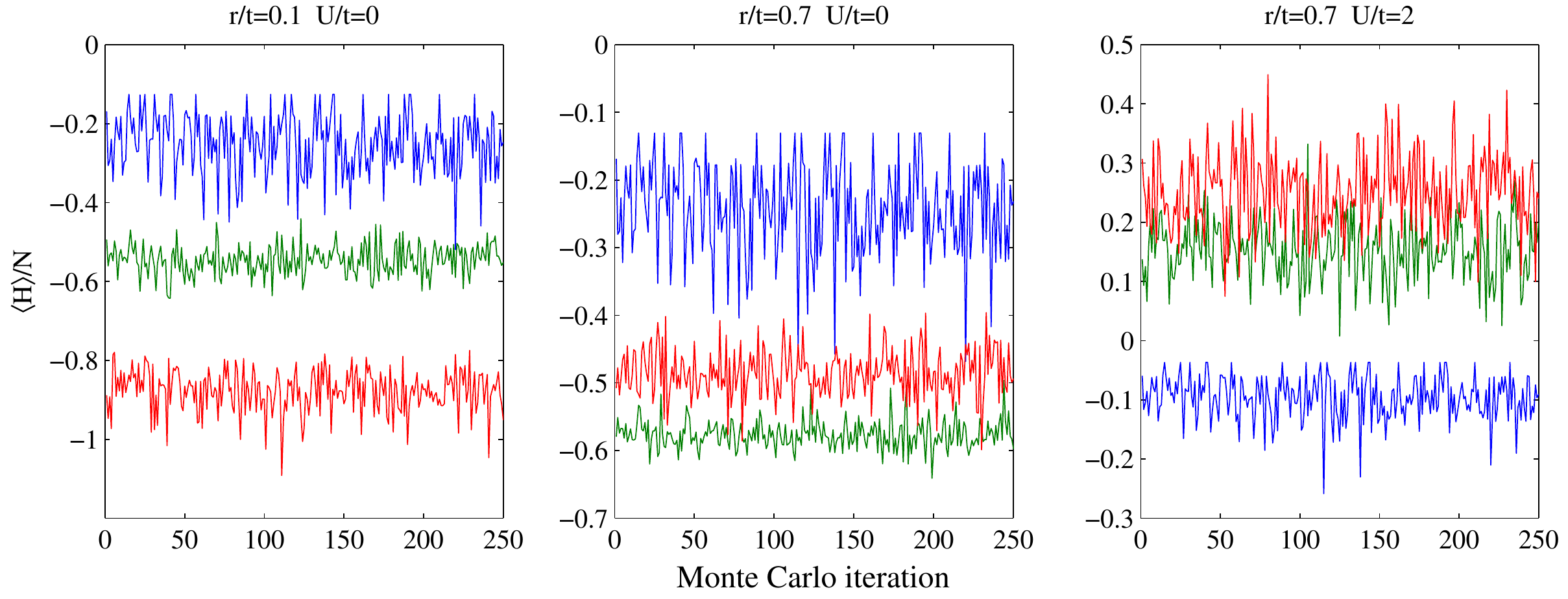}
\caption{\label{fig:var_E}  (Color online) Vartiational energy per site in different regions of the parameter space $(r/t,U/t)$. Colors represent different variational parameters $(\Delta_2,m_2)$: red (SF) $(-1,0)$, green (FCI) $(1,0)$,  and blue (MI) $(1,8)$. From left to right SF, FCI and MI gives the minimum energy respectively.}
\end{figure*}

The variational energy $E=\langle \Psi | H |\Psi \rangle/\langle \Psi | \Psi \rangle$ ($H$ is defined in Eq.~(\ref{eq:BH_model})) is minimized in $(m_2,\Delta_2)$ parameter space of $f_2$. Figure~\ref{fig:var_E} shows this quantity for three representative wave functions in a Monte Carlo run for three different regions of the phase diagram. It is interesting to note that in the middle panel of Fig.~\ref{fig:var_E}, where the FCI phase is expected to stabilize, the FCI candidate wave-function has indeed a relatively small variance. 

\begin{table}
\begin{tabular}{l  c  c c} 
\hline
Transition & Critical point$ \ \ \ $ & $\nu \ \ \ $ & $\eta$  \\
  \hline                       
SF-MI $(t/U)$ & $0.45\pm 0.04 \ \ \ $ & $0.51\pm 0.06 \ \ \ $ & $0.64\pm 0.05$ \\
SF-FCI $(r/t)$ & $0.59\pm 0.02 \ \ \ $ & $0.14\pm 0.02 \ \ \ $ & $0.58\pm 0.08$ \\
  \hline  
\end{tabular}
\caption{\label{tab:cr_exp} The critical exponent $\xi\sim |A-A_c|^{-\nu}$ where $A$ is the tuning parameter shown in parenthesis in front of each transition. The third (unquoted) parameter in each case is fixed: SF-MI $(r=0)$ and SF-FCI $(U=0)$.}
\end{table}

\begin{figure*}
\includegraphics[scale=.9]{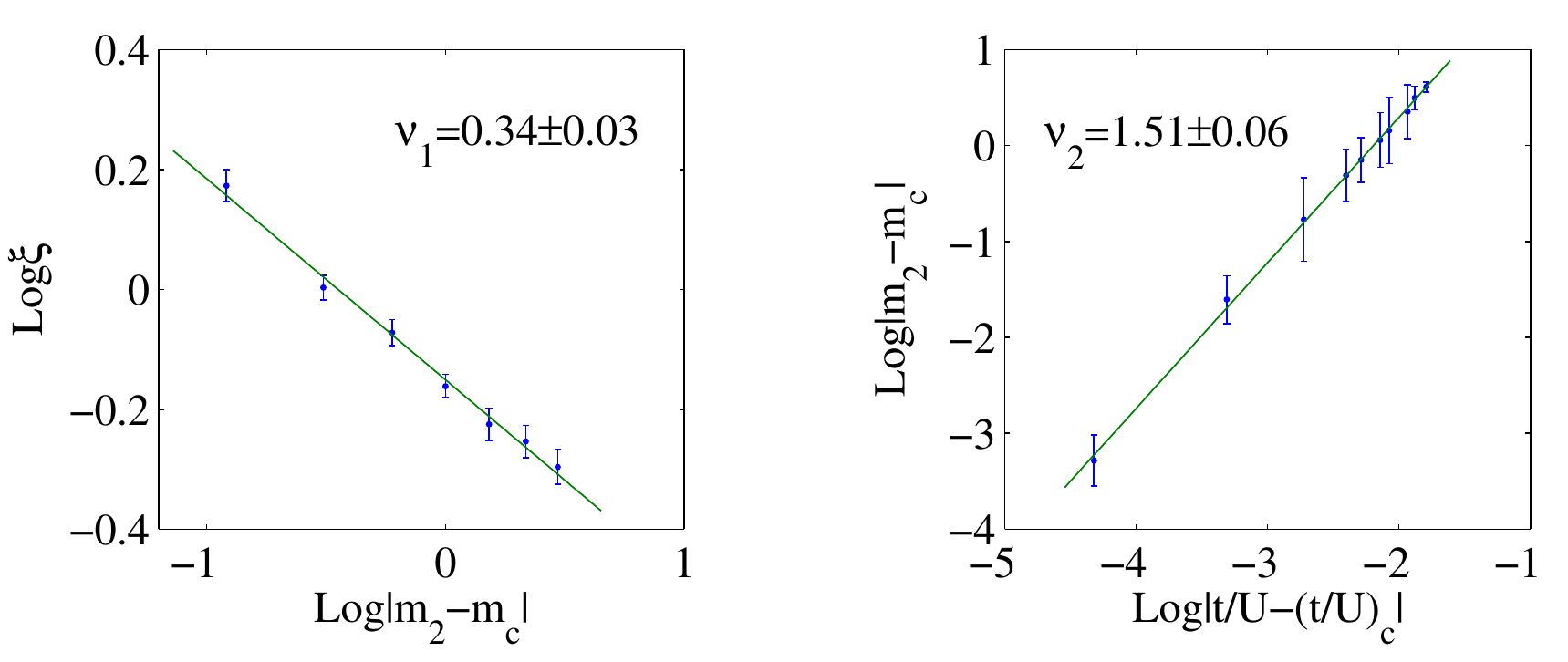}
\includegraphics[scale=.85]{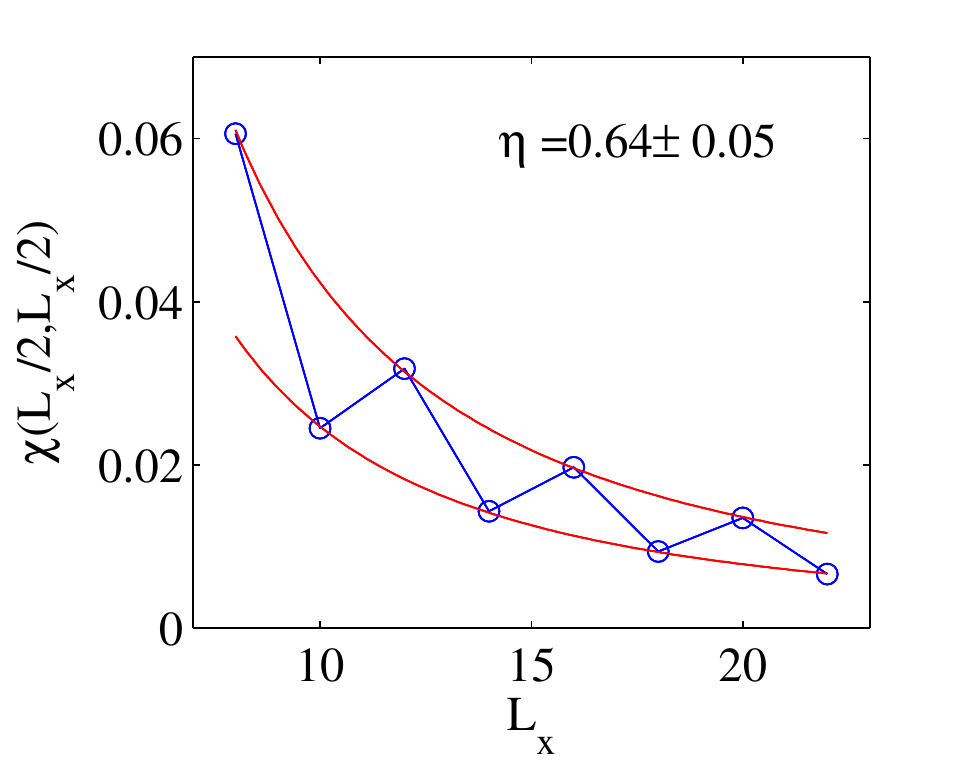}
\caption{\label{fig:SFMI_CR}  (Color online)  The critical exponents for SF-MI transition. Top left: step one is to find $\xi \propto |m_2-m_c|^{\nu_1}$. Top right: step two to find $|m_2-m_c|\propto |t/U- (t/U)_c|^{\nu_2}$. Bottom: finite size scaling of SPDM to derive the anomalous dimension. Errorbars are smaller than symbols. The system size is $16\times 16$.}
\end{figure*}

\begin{figure*}
\includegraphics[scale=.9]{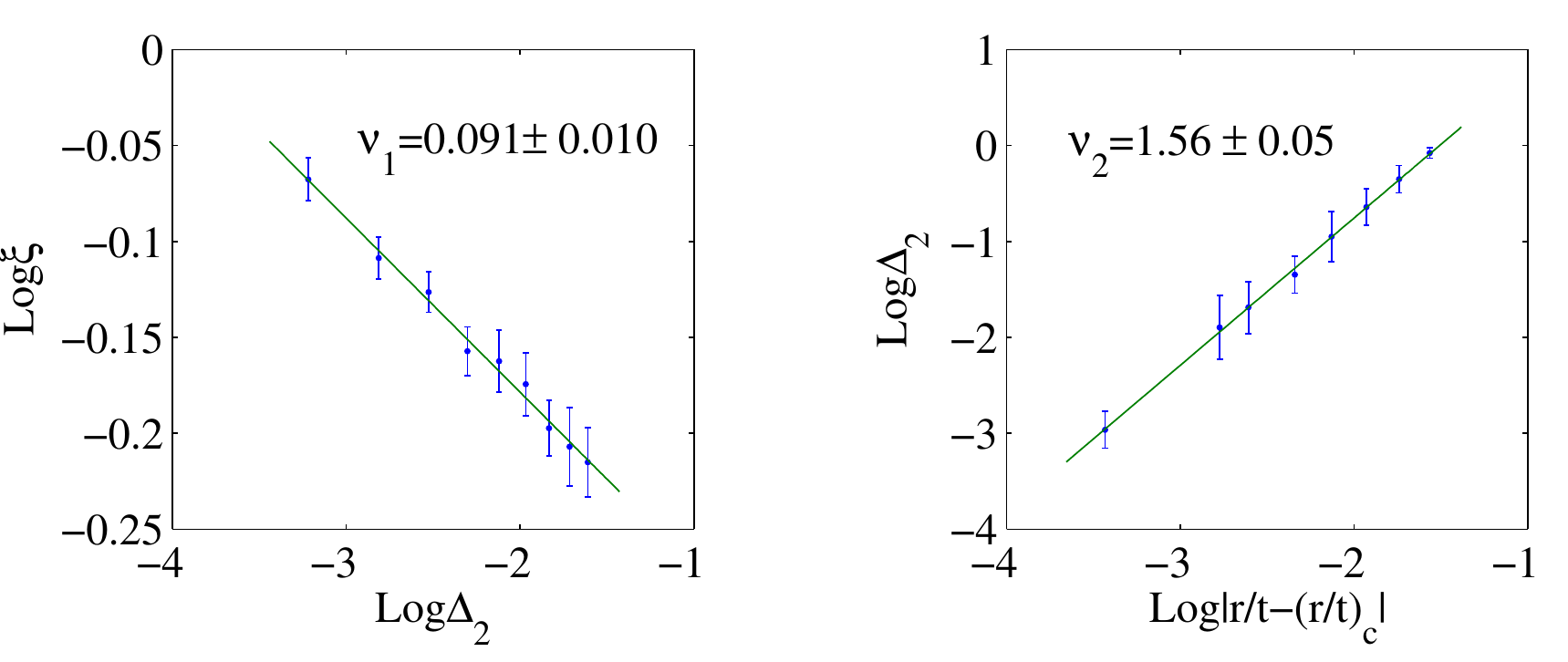}
\includegraphics[scale=.9]{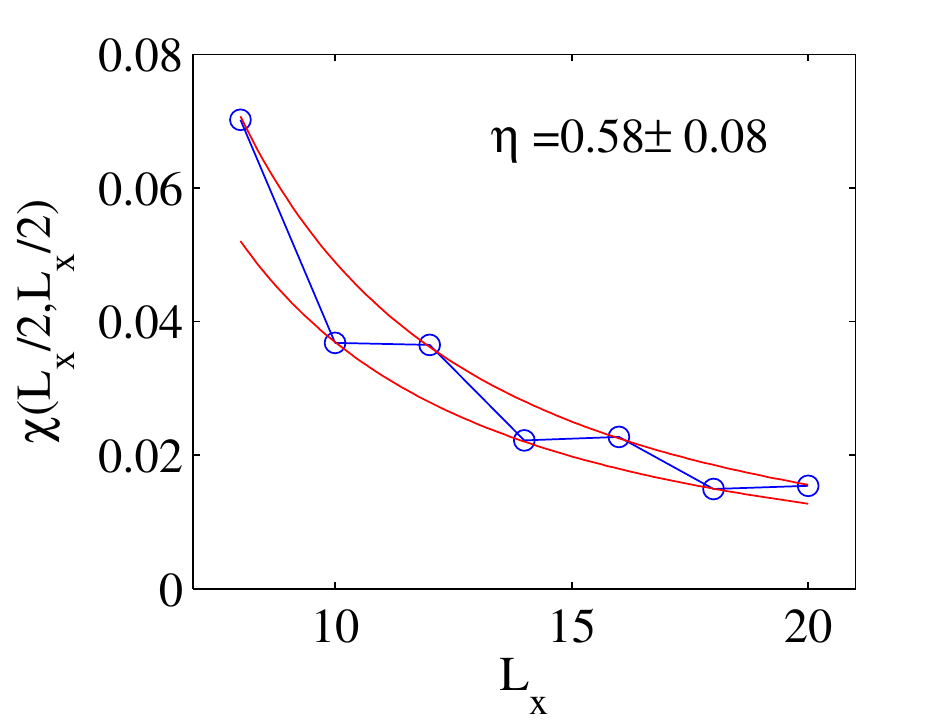}
\caption{\label{fig:FQSF_CR}  (Color online) The critical exponents for SF-FCI transition. Top left: step one is to find $\xi \propto |\Delta_2|^{\nu_1}$. Top right: step two to find $|\Delta_2|\propto |r/t- (r/t)_c|^{\nu_2}$. Bottom: finite size scaling of SPDM to derive the anomalous dimension.  Errorbars are smaller than symbols.  The system size is $16\times 16$.}
\end{figure*}

\begin{figure*}
\includegraphics[scale=.9]{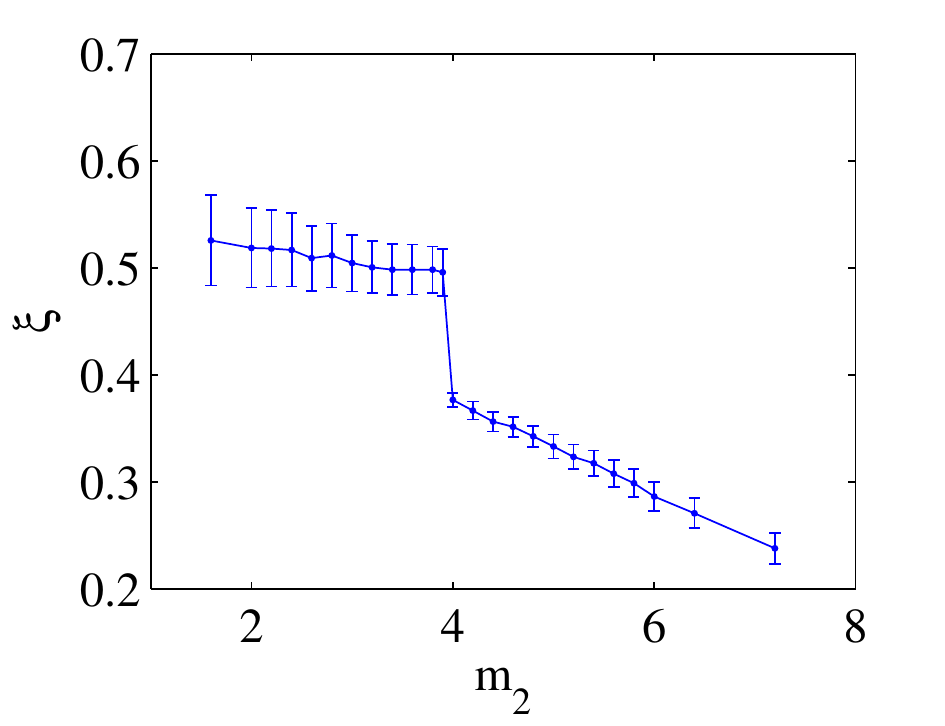}
\caption{\label{fig:FCIMI} The behavior of correlation length across the FCI-MI transition does not show criticality.}
\end{figure*}

Let us now explain how the critical exponents are derived. The results are summarized in Table~\ref{tab:cr_exp}.
 In each transition, we are dealing with two parameters: the parameter within the parton Hamiltonian, call it $A$, and the parameter in the microscopic Hamiltonian, call it $B$. The critical values of these parameters are denoted by $A_c$ and $B_c$ respectively. We extract the critical exponents in two steps: first, we find how the correlation length scales with the parton parameter $\xi\propto |A-A_c|^{\nu_1}$ where $\xi$ is obtained from the best fit of the SPDM. Second, we compute the relation between the two parameters as the transition point is approached  $|A-A_c| \propto |B-B_c|^{\nu_2}$; therefore, the critical exponent is given by $\xi\propto |B-B_c|^\nu$ where $\nu=\nu_1\nu_2$.
Specifically, for SF-MI transition $A=m_2/\Delta_2$ and $B=t/U$ where $r=0$; for SF-FCI transition $A=\Delta_2$ and $B=r/t$ where $U=m_2=0$.
Figures \ref{fig:SFMI_CR} and \ref{fig:FQSF_CR} show the two steps of this procedure. Furthermore, we calculate the anomalous dimension $\eta$ using the finite size scaling of the SPDM (see the lowest plots in Figs.~\ref{fig:SFMI_CR} and \ref{fig:FQSF_CR}). There is a clear even-odd effect in this plot; however, if we fit even and odd data points separately, the exponents will be very close to one another. 
We note that the FCI-MI transition does not show the critical behavior in Fig.~\ref{fig:FCIMI}.

%%%%%%%%%%%%%%%%%%%%%%%%%%%%%%%%%%%%%%%%%%%%%%%%%%%%%%%%
\section{\label{app:szcons} Size consistency of parton wave-functions}

A wave-function is called size consistent if  the total energy $E_{AB}$ is equal to sum of energies of subsystems separately, $E_{AB}=E_A + E_B$, when applied to two uncoupled subsystems $A$ and $B$. This implies that a size consistent wave-function must factor into a product of subsystem wave-functions $|\Psi_{AB}\rangle = |\Psi_A\rangle |\Psi_B\rangle$ as long as the subsystems $A$ and $B$ are not coupled. Building such variational wave-functions for strongly correlated systems has been a long standing issue~\cite{Eric}. Let us illustrate that the projective construction naturally satisfies this condition.
The wave-function for the interacting composite bosons is given by the projected ground state of the non-interacting parton Hamiltonian:
\begin{align} \label{eq:parton_SD}
\Psi (\{\textbf{r}_i\}) = \text{det} \Phi_1(\{\textbf{r}_i\}) \times \text{det} \Phi_2(\{\textbf{r}_i\}),
\end{align}
where $\text{det} \Phi_\alpha(\{\textbf{r}_i\})$ is the Slater determinant for $\alpha$-th parton fermion and the same configuration $\{\textbf{r}_i\}$ for both fermions indicate that they are projected on top of each other. We want to write down a wave-function for $N$ particles in a system consisting of two islands $A$ and $B$ with $N$ sites in each one. As discussed in Appendix~\ref{app:pi-flux}, the parton Hamiltonian is a Chern insulator with a large gap at half filling; therefore, any configuration other than $N/2$ on either sides is going to cost equal or greater than the energy gap and must not be included in the ground state wave-function. So, the ground state wave-function would read
\begin{align}
\Psi_{AB} (\{\textbf{r}^N_i\})= \Psi_{A} (\{\textbf{r}^{N/2}_i\}) \Psi_{B} (\{\textbf{r}^{N/2}_j\}).
\end{align}

An alternative view is as follows. The wave-function can be written explicitly in terms of projected Slater determinants as in Eq.~(\ref{eq:parton_SD}). Here each $\Phi_{\alpha}$ matrix represents the eigenstates of one flavor of parton fermions and consists of $N/2$ eigenstates in the form of $\phi_{\alpha,i}=[\phi_i^A,0]$ and $N/2$ eigenstates in the form of $\bar{\phi}_{\alpha,i}=[0,\phi_i^B]$ where $i=1,\dots,N/2$ runs over eigenstates of the $\alpha$-th fermion and the row vector $[\dots]$ is written in a basis such that the first $N$ entries correspond to sites within the island $A$ and the  last $N$ entries correspond to sites within the island $B$. Now, we can write the general form of wave-function as
\begin{widetext}
\begin{align*}
\Psi_{AB} (\{\textbf{r}^N_i\})&= \sum_n C_n \text{det}\ \left( \begin{array}{ccc}
\phi_{1,i_1} (\textbf{r}_1) &\dots & \phi_{1,i_1} (\textbf{r}_N) \\
\vdots &\vdots &\vdots \\
\phi_{1,i_n} (\textbf{r}_1)& \dots & \phi_{1,i_n} (\textbf{r}_N) \\
\bar{\phi}_{1,i_{n+1}} (\textbf{r}_1)& \dots & \bar{\phi}_{1,i_{n+1}} (\textbf{r}_N)\\
\vdots &\vdots &\vdots \\
\bar{\phi}_{1,i_N} (\textbf{r}_1)& \dots & \bar{\phi}_{1,i_N} (\textbf{r}_N) 
\end{array} \right) 
\times
 \text{det}\ \left( \begin{array}{ccc}
\phi_{2,i_1} (\textbf{r}_1) &\dots & \phi_{2,i_1} (\textbf{r}_N) \\
\vdots &\vdots &\vdots \\
\phi_{2,i_n} (\textbf{r}_1)& \dots & \phi_{2,i_n} (\textbf{r}_N) \\
\bar{\phi}_{2,i_{n+1}} (\textbf{r}_1)& \dots & \bar{\phi}_{2,i_{n+1}} (\textbf{r}_N)\\
\vdots &\vdots &\vdots \\
\bar{\phi}_{2,i_N} (\textbf{r}_1)& \dots & \bar{\phi}_{2,i_N} (\textbf{r}_N) 
\end{array} \right) .
\end{align*}
However, as we argued above, there are only $N/2$ states available in $\phi$ or $\bar{\phi}$; hence any configuration with more  than $N/2$ particles on $A$ or $B$ islands have a repeated row in Slater determinant and so vanishes. Thus, we can write
\begin{align*}
\Psi_{AB} (\{\textbf{r}^N_i\})&= \text{det} \left( \begin{array}{cc}
\Phi_{A,1}(\{\textbf{r}^{N/2}_i\}) & 0\\
0& \Phi_{B,1}(\{\textbf{r}^{N/2}_j\})
\end{array} \right) 
\times
\text{det} \left( \begin{array}{cc}
\Phi_{A,2}(\{\textbf{r}^{N/2}_i\}) & 0\\
0& \Phi_{B,2}(\{\textbf{r}^{N/2}_j\})
\end{array} \right)\\
&=
\left( \text{det} \Phi_{A,1}(\{\textbf{r}^{N/2}_i\})\ \text{det} \Phi_{A,2}(\{\textbf{r}^{N/2}_i\}) \right)
\left( \text{det}  \Phi_{B,1}(\{\textbf{r}^{N/2}_j\})\ \text{det} \Phi_{B,2}(\{\textbf{r}^{N/2}_j\}) \right) \\
&= \Psi_{A} (\{\textbf{r}^{N/2}_i\}) \Psi_{B} (\{\textbf{r}^{N/2}_j\}),
\end{align*}
where we use the shorthand notation $\Phi_{A,\alpha}(\{\textbf{r}^{N/2}_i\})$ and $\Phi_{B,\alpha}(\{\textbf{r}^{N/2}_j\})$ for non-zero elements of the  $\phi_{\alpha,m}(\textbf{r}_n)$ and $\bar{\phi}_{\alpha,p}(\textbf{r}_q)$ matrices respectively. 
\end{widetext}

\end{appendix}

%%%%%%%%%%%%%%%%%%%%%%%%%%%%%%%%%%%%%%%%%%%
\bibliography{PRB_v2}

\end{document}